# Deconstructing the Crystal Ball: From Ad-Hoc Prediction to Principled Startup Evaluation with the SAISE Framework

Seyed Mohammad Ali Jafari[1]

Ali Mobini Dehkordi[2]

Ehsan Chitsaz[3]

Yadollah Yaghoobzadeh[4]

## Abstract

The integration of Artificial Intelligence (AI) into startup evaluation represents a significant technological shift, yet the academic research underpinning this transition remains methodologically fragmented. Existing studies often employ ad-hoc approaches, leading to a body of work with inconsistent definitions of success, atheoretical features, and a lack of rigorous validation. This fragmentation severely limits the comparability, reliability, and practical utility of current predictive models.

To address this critical gap, this paper presents a comprehensive systematic literature review of 57 empirical studies. We deconstruct the current state-of-the-art by systematically mapping the features, algorithms, data sources, and evaluation practices that define the AI-driven startup prediction landscape. Our synthesis reveals a field defined by a central paradox: a strong convergence on a common toolkit—venture databases and tree-based ensembles—but a stark divergence in methodological rigor. We identify four foundational weaknesses: a fragmented definition of "success," a divide between theory-informed and data-driven feature engineering, a chasm between common and best-practice model validation, and a nascent approach to data ethics and explainability.

In response to these findings, our primary contribution is the proposal of the Systematic AI-driven Startup Evaluation (SAISE) Framework. This novel, five-stage prescriptive roadmap is designed to guide researchers from ad-hoc prediction toward principled evaluation. By mandating a coherent, end-to-end methodology that emphasizes stage-aware problem definition, theory-informed data synthesis, principled feature engineering, rigorous validation, and risk-aware interpretation, the SAISE framework provides a new standard for conducting more comparable, robust, and practically relevant research in this rapidly maturing domain.

## Keywords

Startup Success Prediction, Venture Capital, Artificial Intelligence, Machine Learning, Systematic Literature Review.

[1] PhD candidate, Technological Entrepreneurship Department, Faculty of Entrepreneurship, University of Tehran, Iran, sma_jafari@ut.ac.ir
[2] Full Professor, Faculty of Entrepreneurship, University of Tehran, Iran, mobini@ut.ac.ir
[3] Associate Professor, Faculty of Entrepreneurship, University of Tehran, Iran, chitsaz@ut.ac.ir
[4] Assistant Professor, School of Electrical and Computer Engineering, University of Tehran, Tehran, Iran, y.yaghoobzadeh@ut.ac.ir

# 1. Introduction

## 1.1. The High-Stakes Context: The Rise of AI in Venture Capital

The evaluation of startups is a high-stakes endeavor at the heart of the modern innovation economy. Venture capital decisions not only allocate billions of dollars annually but also shape the technological frontier, determining which new ideas receive the resources to grow and potentially redefine industries. Historically, this evaluation process has been characterized as an "art," relying heavily on the intuition, experience, and network of individual investors. However, this traditional paradigm is undergoing a profound transformation. The promise of Artificial Intelligence (AI) and Machine Learning (ML) is compelling, increasingly framed not as a replacement for human judgment but as a powerful augmentation tool (Jarrahi, 2018; Shepherd & Majchrzak, 2022). These technologies offer the potential to mitigate cognitive biases and uncover complex, non-linear patterns in the vast data trails left by modern ventures, thereby improving the efficiency and accuracy of investment decisions.

## 1.2. The Emerging Problem: A Fragmented and Ad-Hoc Methodological Landscape

As the application of AI in this domain has surged, so too has the academic literature seeking to develop and validate predictive models. Yet, despite its vibrancy, the field is characterized by significant methodological fragmentation. An ad-hoc approach persists in much of the research, where critical choices regarding data, features, and validation are often driven by convenience rather than a coherent, integrated strategy.

This fragmentation manifests in several ways. First, the very definition of the outcome variable, "startup success," is inconsistent across studies, ranging from traditional liquidity events like an IPO or M&A (e.g., Setty et al., 2024) to bespoke, expert-rated "sustainability scores" (e.g., Takas et al., 2024). Second, a deep divide persists between data-driven convenience and theory-informed feature engineering; many studies utilize features simply because they are available in a given dataset (e.g., Cholil et al., 2024), while a smaller stream explicitly derives features from established management theories like the Resource-Based View (e.g., Font-Cot et al., 2025). Finally, models are rarely subjected to the same rigorous validation standards, with critical issues like look-ahead bias often being overlooked, a key weakness addressed by only a few methodologically focused studies (e.g., Żbikowski & Antosiuk, 2021). This lack of a common methodological foundation severely limits the comparability, reliability, and ultimately, the practical utility of the resulting models, creating an urgent need for a systematic synthesis of the field.

## 1.3. The Solution: A Systematic Literature Review

To address these critical inconsistencies and provide a clear path forward, this paper conducts a systematic literature review (SLR) of the empirical research on AI-based startup success forecasting. Following a rigorous, pre-defined protocol, we searched leading academic databases, including Scopus and Web of Science, and screened 273 articles to arrive at a final corpus of 57 empirical studies. By systematically deconstructing and synthesizing the methodologies and findings of this corpus, we aim to create the first comprehensive, evidence-based map of this rapidly evolving research domain.

### 1.4. Research Questions and Contributions

This review is guided by five central research questions designed to move from analysis to prescription:

- **RQ1:** What indicators, features, and target variables are commonly used in AI/ML models predicting startup success or failure?
- **RQ2:** What AI/ML techniques have been employed for startup success prediction, and how do they compare?
- **RQ3:** What data sources have been utilized for building AI/ML models in startup evaluation?
- **RQ4:** What are the gaps, limitations, and methodological weaknesses in existing AI/ML-based research for startup evaluation and success prediction?
- **RQ5:** Can a comprehensive conceptual framework be developed to guide the precise selection of data sources, features, and AI/ML models for startup success prediction?

Accordingly, this paper makes two primary contributions to the literature. First, it provides a comprehensive deconstruction of the current state-of-the-art, offering a detailed map of the features, algorithms, data sources, and evaluation practices that define the field. Second, in response to the identified weaknesses, it proposes the Systematic AI-driven Startup Evaluation (SAISE) Framework, a novel, five-stage prescriptive roadmap designed to guide future research toward a more rigorous, coherent, and impactful standard.

### 1.5. Structure of the Paper

The remainder of this paper is organized as follows. Section 2 provides the theoretical background on the relationship between startup evaluation and artificial intelligence. Section 3 describes the systematic review protocol used for study selection. Section 4 presents a detailed thematic analysis of the literature, addressing RQ1, RQ2, and RQ3. Section 5 introduces our primary contribution—the SAISE framework—which answers RQ5. Section 6 discusses the broader implications of our findings, systematically answers RQ4 by detailing the identified gaps, and proposes a concrete agenda for future research. Finally, Section 7 offers concluding remarks.

# 2. The Symbiosis of AI and Startup Evaluation

The evaluation of a startup has traditionally been characterized as an "art," a high-stakes judgment call made under conditions of profound uncertainty. This art relies on an investor's tacit knowledge and the "soft" signals—such as founder personality and team dynamics—that are not easily quantified (Dellermann et al., 2017). However, this paradigm is undergoing a fundamental shift. The integration of Artificial Intelligence (AI) and Machine Learning (ML) does not seek to replace the art of evaluation but rather to augment it, creating a powerful human-machine symbiosis that promises to enhance decision-making in this complex domain (Jarrahi, 2018). This section will conceptually frame the relationship between AI and startup evaluation, establishing the theoretical groundwork for the systematic review that follows.

## 2.1 The Duality of Judgment: Analytical vs. Intuitive Decision-Making

Organizational decision-making, particularly in volatile environments like venture capital, operates on a dual-process model: one process is analytical and rule-based, while the other is intuitive and holistic (Kahneman, 2011). The traditional VC approach has leaned heavily on the latter. Faced with ambiguous information and "unknowable risk"—where representative historical data for a truly novel idea may not exist—investors must rely on their intuition and experience to assess the potential of a venture (Dellermann et al., 2017).

As Jarrahi (2018) notes, this intuitive approach is essential for navigating equivocality and uncertainty. However, it is also susceptible to the cognitive biases inherent in all human judgment. The promise of AI, therefore, is to introduce a powerful analytical counterpart to this process. AI systems excel at the systematic, unbiased processing of vast, "hard" datasets, identifying complex patterns and correlations that are beyond human cognitive capacity (Jarrahi, 2018). This capability is perfectly suited to addressing the complexity of the startup ecosystem, if not its inherent uncertainty.

## 2.2 AI as an Augmentation, Not a Replacement

The narrative of AI in high-stakes decision-making is not one of replacement but of augmentation. The most effective outcomes are consistently achieved not by pitting human against machine, but by combining their complementary strengths (Jarrahi, 2018). This vision of "human-AI symbiosis" is particularly relevant for startup evaluation.

Shepherd & Majchrzak (2022) frame this by arguing that AI can "augment entrepreneurs' [and by extension, evaluators'] ability to notice potential opportunities." The machine can systematically process and connect vast streams of structured data—financial histories, patent filings, market trends—while the human expert remains the ultimate interpreter of "soft" signals like founder personality and team cohesion. This creates a division of cognitive labor: the AI handles the "hard" analytical tasks, freeing up human decision-makers to focus on the intuitive judgments where they hold a comparative advantage. The "centaur" model, where a human-chess-player paired with an AI consistently outperforms either a human or an AI alone, serves as a powerful analogy for this synergistic potential (Jarrahi, 2018).

## 2.3 From Concept to Practice: The Role of Machine Learning

Machine learning (ML) is the engine that operationalizes this AI-driven augmentation. ML algorithms are designed to learn patterns from historical data to make probabilistic predictions about future outcomes (Giuggioli & Pellegrini, 2022). In the context of startup evaluation, this involves training models on large datasets of past ventures to predict the future success of a new one.

This data-driven shift has moved the field from purely theoretical models of success to empirically testable predictive frameworks. As this review will show, researchers now routinely leverage large-scale databases like Crunchbase to build and validate models that can "screen potential investments using publicly available information, diverting [investor] time instead into mentoring and monitoring" (Ross et al., 2021). The most advanced of these applications are even beginning to build risk-aware models, using techniques like cost-sensitive learning to align the model's predictive objective with the asymmetric risk profile of a VC investor, a profound example of augmenting human judgment (Setty et al., 2024).

# 3. Methodology

To perform an accurate and comprehensive analysis of AI-driven startup evaluation, this study adopts a systematic literature review (SLR) methodology, following the principles outlined by Tranfield et al. (2003). This approach allows for a replicable, transparent, and rigorous synthesis of the existing body of research.

### 3.1. Search Strategy and Databases

The search process was designed to identify all relevant empirical studies at the intersection of startup evaluation and artificial intelligence. We conducted our search in June, 2025 across two leading academic databases, Scopus and Web of Science, chosen for their extensive coverage of peer-reviewed literature in the relevant subject areas.

The final search string was constructed from four core conceptual blocks to ensure high precision and recall:

```
( "forecasting" OR "prediction" OR "classification" OR "prognosis" OR
"evaluation" ) AND ( "startup" OR "start-up" OR "new venture" OR
"entrepreneurship" ) AND ( "success" OR "survival" OR "failure" OR
"performance" ) AND ("artificial intelligence" OR "ai" OR "machine
learning" OR "ml" OR "deep learning")
```

This query was applied to the title, abstract, and keywords of articles in the databases, with no starting date restriction to capture the full history of the field. all keyword searches were conducted on a case-insensitive basis.

### 3.2. Inclusion and Exclusion Criteria

To define the scope of the review and ensure the relevance of the final corpus, we established a strict set of inclusion and exclusion criteria. The review was limited to peer-reviewed journal articles and conference proceedings published in English. A study was included if it met all of the following conditions:

- It presented an empirical study involving data analysis.
- It developed or tested a computational model using AI, machine learning, or related data science techniques.
- Its prediction target was related to firm-level success (e.g., survival, exit, funding, performance).

A study was excluded if it was a non-empirical work (e.g., literature review, editorial, conceptual-only paper), if its primary focus was not on firm-level outcomes (e.g., predicting crowdfunding campaign success or student entrepreneurial intent), or if it did not use AI/ML techniques for prediction.

A key decision in our protocol was to include both peer-reviewed journal articles and conference proceedings. This choice was made deliberately to reflect the dynamic nature of the Artificial Intelligence field. Unlike many other disciplines where journals are the sole primary outlet for significant findings, in computer science and AI, premier conferences (e.g., NeurIPS, ICML, KDD, CIKM) serve as critical venues for the dissemination of groundbreaking, peer-reviewed research. Due to the rapid pace of innovation, the faster publication cycle of conferences means that many novel algorithms, datasets, and methodological approaches appear in this format first. To exclude conference papers would be to ignore a substantial portion of the most current and innovative work, thereby creating an incomplete and outdated picture of the field. Our approach ensures that this review captures the full spectrum of high-quality, impactful research.

### 3.3. Study Selection and Screening Process

The study selection followed the multi-stage PRISMA 2020 guidelines for systematic reviews (Page et al., 2021). The process is visually summarized in the PRISMA flow diagram in Figure 1. The identification phase began with a search across the Scopus and Web of Science databases, which yielded a total of 273 records. In the screening phase, 63 duplicate records were removed, leaving 210 unique articles. The titles and abstracts of these articles were then screened against the inclusion criteria, which resulted in the exclusion of 125 articles that were clearly not focused on AI-based startup success prediction. This left 85 articles for the eligibility phase, which involved a full-text review. At this stage, a further 28 articles were excluded for being non-empirical, having an irrelevant outcome variable, or not using an AI/ML methodology. This rigorous process resulted in a final corpus of 57 empirical studies that were included in the thematic synthesis.

**Figure 1**. PRISMA 2020 flow diagram illustrating the systematic study selection process. The diagram outlines the number of records at each stage of the review: identification, screening, and eligibility, resulting in the final corpus of 57 empirical studies included in the thematic synthesis.

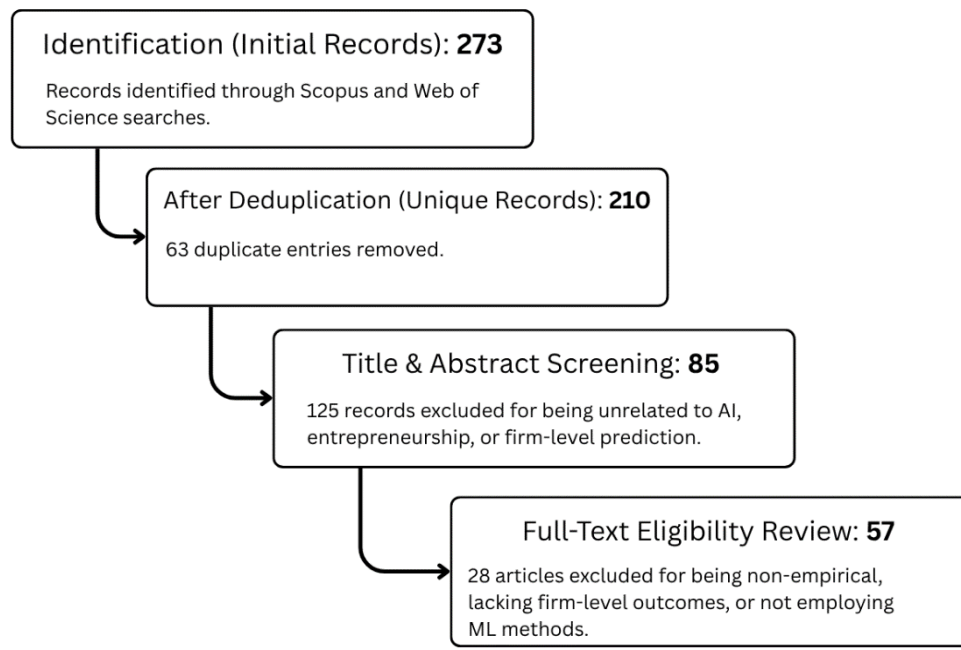

### 3.4. Data Extraction and Thematic Synthesis

For each of studies in the final corpus, we systematically extracted key data points to address our research questions. This included all predictor variables used, the operational definition of the target variable, the AI/ML algorithms tested, the validation and tuning methods employed, the use of XAI techniques, and the primary data sources. Following this extraction, a thematic synthesis was conducted to identify, categorize, and quantify dominant patterns, methodological gaps, and emerging frontiers across the literature, the results of which are detailed in Section 4.

# 4. Results

This section systematically presents the synthesized findings from the 57 empirical studies that constitute our final corpus. Our analysis reveals a field defined by a central paradox: a strong convergence on a common toolkit—venture databases, specific feature families, and tree-based algorithms—but a stark divergence in methodological rigor and definitional consistency. We structure this section to first deconstruct the points of convergence before exploring the critical areas of divergence. We begin by analyzing the features and target variables that define the prediction landscape (RQ1), then dissect the algorithms and validation practices (RQ2), and finally, inventory the underlying data sources (RQ3).

# 4.1. The Anatomy of Prediction: Features, Indicators, and Target Variables (RQ1)

Our analysis of the predictors and outcomes used in AI/ML startup evaluation begins with the features themselves (RQ1). Here, we uncover the first layer of our central paradox: a clear convergence on *what* features to use, but a significant divergence in *how* those features are justified and, most critically, in how the 'success' they predict is even defined. We structure this analysis in three parts: first, mapping the consensual feature landscape; second, moving from feature frequency to predictive impact; and finally, deconstructing the fragmented definitions of 'startup success' used as model targets.

### 4.1.1. The Feature Landscape: A Methodological Divide Between Theory and Data-Driven Approaches

The convergence of the field is immediately apparent in the feature landscape. Our exhaustive cataloging of predictor variables reveals a strong and consistent consensus on a core "money-people-market" triad. Table 1 presents a thematic overview of these feature families with representative studies, while Appendix A contains the exhaustive list. However, this surface-level agreement masks a foundational divergence in how these features are selected. The dominant approach is data-driven, utilizing features primarily because they are conveniently present in a source (e.g., Cholil et al., 2024). In stark contrast, a smaller but methodologically distinct set of studies adopts a theory-driven approach, grounding their feature selection in established management theories like the Resource-Based View (e.g., Font-Cot et al., 2025). Table 1 provides a comprehensive thematic analysis of all feature families.

Many papers, particularly those built on readily available datasets like Kaggle, adopt a data-driven approach, utilizing features primarily because they are present in the source (e.g., Cholil et al., 2024). In contrast, a smaller but methodologically distinct set of studies first grounds their feature selection in established management and entrepreneurship theories. For instance, Font-Cot et al. (2025) explicitly derive their features from theories like the Resource-Based View (RBV) and Anti-Fragility, a practice that lends strong theoretical validity to their model but often necessitates primary data collection via questionnaires. This highlights a critical challenge for the field: balancing theoretical rigor with the practical limitations of large-scale data sources.

Despite these differing philosophies, the resulting feature landscape is remarkably consistent. The most prevalent families are Funding History (39/57 studies, or 68%), followed by Team/Founder Attributes (34/57, or 60%) and Market/Sector Tags (34/57, or 60%). However, a crucial insight emerges from our analysis: the reliance on these features, especially financial ones, creates a significant "pre-seed prediction problem." As Razaghzadeh Bidgoli et al. (2024) implicitly show, models that lean heavily on funding data are inherently limited to evaluating startups that have already passed the seed stage. This is a point of friction, given that a survey of VCs by Takas et al. (2024) revealed a strong demand for AI systems capable of evaluating precisely these pre-seed ventures, for which financial histories are non-existent.

**Table 1: Thematic Landscape and Prevalence of Predictor Feature Families (Abridged, N=57)**

This table provides a hierarchical overview of the feature families identified across the 57-study corpus, ranked by their frequency of use. It highlights the field's strong consensus on a core "money-people-market" triad while also cataloging the long tail of more specialized and emerging indicators.

| Rank | Feature Family | Core Concept & Representative Variables | Prevalence (N=57) | Representative Studies (Citations) |
|---|---|---|---|---|
| 1 | Funding History & Cadence | The magnitude, timing, and velocity of capital acquisition. Total funding (USD), # of rounds, time-to-first-round, gap-between-rounds. | 39 (68%) | Setty et al. (2024), Ross et al. (2021), Belgaum et al. (2024), Gangwani et al. (2023) |
| 2 | Team / Founder Attributes | The human capital, experience, and composition of the founding team. Founder count, prior exits, education pedigree, skill diversity. | 34 (60%) | Font-Cot et al. (2025), Li, Y. et al. (2024), Dellermann et al. (2017), Pasayat & Bhowmick (2023) |
| 3 | Market / Sector Tags | The industrial and competitive context in which the startup operates. One-hot industry codes, BERT topic vectors, market growth dummies. | 34 (60%) | Kim et al. (2023), Żbikowski & Antosiuk (2021), Chen. (2024), Guerzoni et al. (2021) |
| 4 | Investor Structure / Quality | The quality and composition of the startup's financial backers. Investor count, lead/brand flags, top-tier investor dummy. | 26 (46%) | Guerzoni et al. (2021), Setty et al. (2024), Pasayat & Bhowmick (2021), Abhinand & Poonam (2022) |
| 5 | Digital & Social Traction | Proxies for market presence, brand momentum, and user engagement. Website liveness, LinkedIn/Twitter followers, web mention freshness. | 20 (35%) | Razaghzadeh Bidgoli et al. (2024), Te, Liu, & Müller (2023), Sharchilev et al. (2018), Ross et al. (2021) |
| 6 | Product / Technology Sophistication | The maturity and defensibility of the startup's core technology or product. Patent counts, Technology Readiness Level (TRL), GitHub velocity. | 14 (24%) | Ross et al. (2021), Rawat et al. (2025), Dellermann et al. (2017), Ferrati et al. (2021) |
| 7 | Financial-Statement Ratios | Traditional accounting metrics reflecting a firm's financial health. Cash/asset ratio, debt-to-cash-flow, EBITDA margin, burn-rate. | 8 (14%) | Liu et al. (2022), Adebiyi et al. (2024), Guerzoni et al. (2021), Font-Cot et al. (2025) |
| 8 | Sentiment & Emotion | Measures of public or investor affect derived from textual data. VADER polarity, ROBERTa emotion vectors, tweet sentiment. | 3 (5%) | Antretter et al. (2018), Goossens et al. (2023), Sharchilev et al. (2018) |
| 9 | Psychometric / Behavioural | Direct measures of founder personality and behavioral traits. Big-Five OCEAN scores, need-for-achievement, resilience. | 4 (7%) | Takas et al. (2024), Font-Cot et al. (2025), Sabahi & Parast (2020) |
| 10 | ESG & Sustainability Cues | Indicators reflecting a startup's focus on environmental or social goals. E/S/G keyword dummies, green-impact flags. | 2 (3%) | Takas et al. (2024), Rani et al. (2024) |
| 11 | Novel Engineered / Interaction Metrics | Composite features modeling the interplay between different entities. Investor-Market-Fit, Founder-Idea-Fit cosine. | 3 (5%) | Gangwani et al. (2023), Li, Y. et al. (2024), Ross et al. (2021) |

*Note: This is an abridged version showing representative studies. For an exhaustive list, please see the full table in Appendix A.*

### 4.1.2. From Static Features to Dynamic Signals: Context, Momentum, and the Research Frontier

Moving from feature frequency to predictive impact reveals a more nuanced and insightful picture. While Table 1 shows what features are commonly *used*, it does not tell us what features actually *work*. To answer this, we synthesized the findings exclusively from those studies in our corpus that went a step further, employing post-hoc analytical techniques to interrogate their models and determine which features were most influential in driving predictions. These methods, such as SHAP, MDI, and feature ablation, provide crucial insights into the model's decision-making process that are not apparent from performance metrics alone.

The results of this synthesis are presented in Table 2, which organizes these high-impact predictors into emergent thematic clusters. The analysis confirms a central argument of our review: dynamic signals that capture momentum, external validation, and relational context consistently provide greater predictive lift than static, firm-level attributes.

This trend is evident across several key themes. For instance, in the Social & Digital Traction cluster, Razaghzadeh Bidgoli et al. (2024) used SHAP analysis to identify LinkedIn/Twitter followers as top drivers, representing a live measure of market engagement. Similarly, in the Relational & Contextual Factors cluster, the work of Li et al. (2024) stands out by demonstrating that Cohort network features—which measure a startup's embeddedness in its peer ecosystem—were the most powerful predictors in their model. This contrasts sharply with more traditional, static features like industry classification, which, while frequently used, rarely emerge as top-ranked predictors in these deep-dive analyses.

This evidence strongly suggests a maturation in the field. The focus is shifting from *what a startup is* (its static profile) to *what a startup is doing* (its dynamic signals) and *where it sits* (its relational context). This fact-based synthesis, grounded in the specific analytical findings of the source papers, provides a much clearer picture of the true drivers of predictive power in startup success modeling.

This need for dynamic, contextual signals is driving the research frontier toward the engineering of novel, often relational, indicators (Table 3). Here, the most innovative studies are moving beyond what is available in standard databases to create proprietary signals. The work of Żbikowski & Antosiuk (2021) stands out by using features that are intentionally robust against look-ahead bias, avoiding funding data that would not be available at the point of prediction. This methodological rigor is rare but essential for building practically useful models. Other innovations found across our corpus include the PCA-based dimensionality reduction used by Choi (2024) to manage feature correlation and enhance performance, and the explicit modeling of missing information as a predictive penalty by Ross et al. (2021). These sophisticated approaches signal a maturation of the field, moving beyond simple feature inclusion to thoughtful feature engineering and selection, often using techniques like GreedyStepwise (Vasquez et al., 2023).

**Table 2: Thematic Synthesis of High-Impact Predictors Based on Feature Importance Analysis**

This table synthesizes the most impactful predictor variables as identified by studies in the corpus that employed rigorous post-hoc feature importance techniques (e.g., SHAP, MDI, permutation importance, decision tree splits). Features are grouped into emergent thematic clusters to reveal which types of signals are most influential in modern startup success prediction models.

| Thematic Cluster | Specific Predictive Feature(s) | Core Concept / Interpretation | Supporting Studies & Method |
|---|---|---|---|
| I. Digital Traction & Momentum | LinkedIn/Twitter followers, Website traffic, Time-to-first-investment, Startup age | Proxies for real-time market-facing momentum, brand recognition, and user engagement. Higher values and shorter times-to-funding indicate strong social proof and public interest. | Razaghzadeh Bidgoli et al. (2024) [RF/SHAP]Ross et al. (2021) [XGBoost SHAP]Sharchilev et al. (2018) [Gradient Boosting]Te, Liu, & Müller (2023) [RF/SHAP] |
| II. External Validation: Capital & Reputation | Last raised amount, Total funding, Unique-investor count, Top-tier investor share | Signals of validation from sophisticated capital sources. Represents not just the amount of money, but the *quality, breadth, and reputation* of the investors providing it. | Cholil et al. (2024) [Feature Importance]Abhinand & Poonam (2022) [RF Splits]Vasquez et al. (2023) [GreedyStepwise] |
| III. Team & Human Capital | Number of employees on LinkedIn, Founder's previous experience, Max founder connections, Big-Five Personality Game Scores | Measures of team growth, founder's social & human capital, and intrinsic psychological traits. Moves beyond simple founder counts to dynamic growth and deep behavioral attributes. | Razaghzadeh Bidgoli et al. (2024) [RF/SHAP]Takas et al. (2024) [RF / LOOCV]Te, Liu, & Müller (2023) [RF/SHAP] |
| IV. Core Financial Health | Debt-to-cash-flow ratio, Cash-to-total-assets ratio, 'Is Profitable' (binary) | Traditional, fundamental metrics of solvency and financial stability, especially crucial for predicting distress or survival in later-stage or post-crisis scenarios. | Liu et al. (2022) [XGBoost Splits]Takas et al. (2024) [Decision Tree] |
| V. Relational & Contextual Factors | Cohort network features (e.g., Portfolio Similarity), Investor-Market-Fit Interaction Index | High-order features capturing the firm's embeddedness within its ecosystem. Measures peer effects and the strategic alignment between investors and market trends. | Li et al. (2024) [RF Importance]Gangwani et al. (2023) [Ensemble Methods] |
| VI. Product & Intellectual Property | Longitudinal Patent-Claim Curves, Number of patents, "Scalability" keywords from pitch decks | Signals derived from the core product or IP. Captures the trajectory and defensibility of the innovation itself, moving beyond static patent counts to qualitative or time-series analysis. | Ferrati et al. (2021) [Deep Seq. / Time-series]Samudra & Satya (2024) [LLM-based]Ross et al. (2021) [XGBoost SHAP] |

**Table 3: Catalog of Novel and Rare Indicators at the Research Frontier**

This table inventories the pioneering, infrequently used indicators that signal future research directions. These features, often derived from unstructured data, behavioral science, or sophisticated relational modeling, showcase a clear trajectory toward engineering more nuanced, theory-driven, and context-aware predictors beyond what is readily available in standard databases.

| Thematic Frontier | Novel Indicator / Concept | First (or only) Study in Corpus | Why it is Novel in this Field |
|---|---|---|---|
| Unstructured Data Parsing | GPT-4 Deck Semantics | Samudra & Satya (2024) | First automated extraction of qualitative due-diligence cues (e.g., "scalability" keywords) directly from unstructured pitch-deck PDFs. |
| | Longitudinal Patent-Claim Curves | Ferrati et al. (2021) | Treats intellectual property accumulation as a time-series for a deep learning model, capturing the trajectory of innovation rather than a static count. |
| Relational & Network Modeling | Relational Cohort Features | Li et al. (2024) | Moves the unit of analysis beyond the individual firm to model the "peer effects" and network dynamics within an accelerator cohort. |
| | Investor-Market-Fit Interaction Index | Gangwani et al. (2023) | Engineers a high-order feature that explicitly encodes the alignment between an investor's focus and market momentum, going beyond simple counts. |
| Behavioral & Psychometric Integration | Big-Five Personality Game Scores | Takas et al. (2024) | Injects individual founder psychometrics (e.g., OCEAN traits) harvested from a serious game directly into a success prediction model. |

### 4.1.3. Operationalizing "Success": A Fragmented Landscape of Definitions and Methods

If the divide in feature selection is a crack in the foundation, the lack of a standardized target variable is a fault line running through the entire field. Our analysis confirms that the most critical point of divergence—and arguably the single greatest challenge to comparability and synthesis—is the fragmented operationalization of 'startup success' itself. As detailed in Table 4, which presents an abridged list, the ground truth that models are trained on is inconsistent, ranging from short-term funding events to long-term survival (see Appendix B for the exhaustive list of studies). This makes it impossible to directly compare performance across studies. As we recommend in our discussion, the field would benefit immensely from establishing clearer definitional standards. The landscape is dominated by two primary conceptions of success: Exit Event / Firm Status (19/57 studies) and Funding-Milestone Achievement (13/57 studies). While these are pragmatic and data-driven definitions, they are not without issue: a startup can achieve a Series-A round and still fail, or remain private and be highly successful.

This definitional variety is further compounded by critical methodological challenges inherent in the data. A recurring theme is the problem of imbalanced datasets, where failed startups vastly outnumber successful ones. To prevent models from simply defaulting to the majority class, researchers widely employ techniques like SMOTE (e.g., Liu et al., 2022; Gautam & Wattanapongsakorn, 2024) or strategic under/over-sampling (Vasquez et al., 2023) to create more balanced training distributions. While this improves model accuracy on the minority class, it is a deliberate manipulation of the real-world distribution.

Recognizing these issues, some researchers are pioneering more nuanced targets. Takas et al. (2024) introduce the concept of "startup sustainability," distinguishing it from short-term success by focusing on long-term viability—a concept they operationalize through a multi-level ordinal score. Font-Cot et al. (2025) similarly model a continuous "survival score."

Perhaps the most sophisticated response to these challenges is not to rebalance the data, but to redefine the learning objective itself through cost-sensitive relabeling, as introduced by Setty et al. (2024). Their use of the MetaCost algorithm recalculates the ground-truth labels to explicitly minimize the expected financial risk for an investor, who fears a false positive far more than a false negative. This shifts the modeling objective from mere classification accuracy to risk-optimized decision support, a profound and practical innovation.

Finally, the temporal context of the target variable matters immensely. Liu et al. (2022) brilliantly demonstrate this by splitting their analysis pre- and post-the 2008 financial crisis, showing that the predictive importance of features like cash flow indicators changed dramatically under macroeconomic stress. This finding serves as a powerful reminder that the drivers of startup success are not static but are contingent on the broader economic environment.

### Table 4: Thematic Landscape of Target Variable Operationalizations for "Startup Success" (Abridged, N=57)
This table deconstructs the various ways "startup success" is defined and measured across the reviewed literature (N=57), grouping them into six conceptual themes. It highlights the critical fragmentation of the ground truth.

| Conceptual Theme | Prevalence (N=57) | Operationalization & Representative Examples | Typical Horizon | Core Rationale / Purpose | Representative Studies (Citations) |
|---|---|---|---|---|---|
| Exit Events / Firm Status | 19 (33%) | Success is a firm's ultimate liquidity event or operational status, using binary or multi-class labels. • IPO or M&A (binary exit) • Operating/Acquired/IPO/Closed. | Snapshot to 10 years | Captures the canonical goal of venture capital by identifying a definitive realization of success or failure. | Setty et al. (2024), Ross et al. (2021), Kim et al. (2023), Guerzoni et al. (2021) |
| Funding Milestone | 13 (23%) | Success is a capital acquisition event, typically an institutional VC round. • Raises a first Series-A round • Secures Series-B funding. | 1-5 years | Tracks whether the venture succeeds in attracting external capital, signaling early market validation and investor confidence. | Te, Liu, & Müller (2023), Żbikowski & Antosiuk (2021), Li, Y. et al. (2024), Belgaum et al. (2024) |
| Survival, Lifespan & Sustainability | 6 (11%) | Success is long-term viability, moving beyond discrete events. • 5-year survival probability • Expert-rated sustainability score. | 5 years to full lifespan | Measures the firm's capacity to operate viably over time, often incorporating non-financial aspects. | Font-Cot et al. (2025), Takas et al. (2024), Wang et al. (2022) |
| Entrepreneurial / Project Action | 6 (11%) | Success is defined by the completion of an entrepreneurial action or | Short-term | Focuses on predicting intermediate actions or project-level success rather | Goossens et al. (2023), Rawat et al. (2025), Sabahi & Parast (2020) |

| | | project, distinct from firm-level outcomes. • Predicting investor decisions • Kickstarter campaign success • Shark Tank deal prediction. | | than the venture's ultimate fate. | |
| Valuation & Investor Return | 4 (7%) | Success is quantified via direct economic value. • Market-cap regression • Unicorn status (≥ $1B valuation). | Snapshot to full fund life | Quantifies the economic upside for investors, moving beyond simple binary success. | Chen. (2024), Hu et al. (2022), Abhinand & Poonam (2022) |
| Financial Distress / Health | 2 (4%) | Success is framed as the avoidance of failure or near-term solvency. • Bankruptcy or acquisition within 12 months. | 1-year projection | Provides early-warning signals of impending insolvency, allowing for pre-emptive intervention. | Liu et al. (2022), Adebiyi et al. (2024) |

*Note: This is an abridged version showing representative studies. For an exhaustive list of studies in each category, please see the full table in Appendix B.*

# 4.2. The Modeler's Toolkit: A Landscape of Algorithms, Validation, and Explainability (RQ2)

Beyond the features used, our second research question (RQ2) investigates *how* those features are modeled. This analysis of the modeler's toolkit reveals the paper's central paradox in its starkest form. We find a remarkable convergence on a specific class of algorithms, establishing a de facto industry standard. However, we then uncover a profound divergence in the rigor applied to the rest of the pipeline: model validation, hyper-parameter tuning, data engineering, and explainability. It is this chasm between common practice and best practice that severely limits the reliability and comparability of many studies.

### 4.2.1. Algorithmic Choices: The Unquestioned Reign of Tree-Based Ensembles

As Table 5 demonstrates, the methodological convergence of the field is nowhere more evident than in the choice of algorithms. An abridged version of this table is presented below, with the exhaustive list of citations available in Appendix C. The landscape is unequivocally dominated by tree-based and ensemble methods, with 72% of all studies utilizing at least one tree-based algorithm (e.g., Random Forest) and 67% using an ensemble method (e.g., XGBoost). Their popularity stems from their high performance on the structured, tabular data that characterizes this domain, as well as their built-in mechanisms for feature importance ranking (Cholil et al., 2024; Park et al., 2024). While traditional classifiers like Linear/GLM models (54% of studies) and Kernel/Instance-based methods (53%) remain indispensable as robust and interpretable baselines (e.g., Żbikowski & Antosiuk, 2021), the most potent models are consistently tree-based ensembles. Gradient-boosted frameworks like XGBoost and LightGBM are the de facto headliners in a majority of comparative studies, prized for their ability to capture complex, non-linear interactions and deliver top-tier accuracy (Gautam & Wattanapongsakorn, 2024; Samudra & Satya, 2024).

In contrast, Deep Learning is emerging but not yet dominant, appearing in 35% of the reviewed papers. This includes innovative applications like the Bi-LSTM with custom activation for performance prediction (Srivani et al., 2023) and the hybrid CNN-LSTM for time-series valuation (Chen, 2024). However, the significant data requirements of advanced architectures mean their use is still limited, aligning with the field's general lack of massive, publicly labeled corpora. Finally, more niche approaches like Bayesian Networks (Gujarathi et al., 2024) represent important, though less common, diversifications of the field's methodological portfolio. This hierarchy of algorithmic choice, however, is only part of the story; the rigor of their validation is what ultimately determines the trustworthiness of the findings.

### 4.2.2. Validation and Tuning: A Gap Between Common Practice and Best Practice

While there is clear consensus on which algorithms to use, there is little agreement on how to validate them. This is where the methodological divergence becomes most acute. Our analysis reveals a persistent and troubling gap between common practice and best practice in both model validation and hyper-parameter tuning, undermining the trustworthiness of many published results.

Regarding validation protocols, a simple train/test hold-out split remains a common approach, used in a substantial number of studies (e.g., Zhang & Lau, 2023; Belgaum et al., 2024; Shi et al., 2024). While straightforward, this method is susceptible to sampling bias and is generally insufficient for rigorous academic claims. A growing number of studies now employ more robust k-fold cross-validation (CV), particularly when sample sizes permit (Gautam & Wattanapongsakorn, 2024; Gangwani et al., 2023).

However, a critical observation is the rarity of more advanced validation techniques essential for preventing optimistic bias. For instance, correctly nested CV, which is crucial when performing hyper-parameter tuning, remains exceptionally rare. The work of Te et al. (2023b) stands as a key exemplar of this best practice. Furthermore, only a minority of studies explicitly address look-ahead bias through strict temporal or rolling-origin validation designs (e.g., Wang et al., 2022; Żbikowski & Antosiuk, 2021), a critical safeguard for any model intended for real-world forecasting. The persistence of these less-rigorous validation methods across a large portion of the literature suggests a critical area for collective improvement.

The gap is even more pronounced in hyper-parameter tuning. A striking 50% of studies in our sample either rely on library defaults or perform only minimal manual adjustments (e.g., Kuznietsova & Grushko, 2019; Abhinand & Poonam, 2022). This "minimalist mindset" severely limits the potential performance of the selected algorithms and makes robust model comparisons nearly impossible. When tuning is performed, it is often a shallow, non-systematic search. Formal, automated search strategies like grid search (Gautam & Wattanapongsakorn, 2024), random search (Te et al., 2023b), or meta-heuristic optimization (Hu et al., 2022) are the exception, not the rule. This reveals that even the best algorithm is ineffective without sound validation and tuning. However, before any model is trained, the data itself must be engineered to address its intrinsic challenges.

### 4.2.3. Data Engineering: Taming Imbalance, Dimensionality, and Leakage

This divergence in rigor extends to the foundational stage of data engineering. Before an algorithm can be trained, researchers must address three recurring challenges in the raw data: class imbalance, dimensionality, and information leakage. How they do so is another key differentiator between robust and fragile models. Our review identifies three recurring data challenges that researchers must address: class imbalance, dimensionality, and information leakage.

The most pervasive issue across the corpus is extreme class imbalance, where failed startups vastly outnumber successful ones. To combat this, techniques like SMOTE (Synthetic Minority Over-sampling Technique) are now routinely applied to create more balanced training sets and improve recall on the minority class, a practice seen in numerous studies (e.g., Gautam & Wattanapongsakorn, 2024; Liu et al., 2022; Srivani et al., 2023). Another common challenge, particularly in small-sample or feature-rich studies, is high dimensionality and multicollinearity. Here, researchers are increasingly turning to dimensionality reduction, most notably via Principal Component Analysis (PCA), which can decorrelate predictors and improve the performance and stability of subsequent models (e.g., Choi, 2024; Gracy et al., 2024).

Finally, and most critically, is the challenge of data and information leakage. The most methodologically rigorous studies now explicitly design their feature engineering pipelines to prevent look-ahead bias. The work by Żbikowski & Antosiuk (2021) and Li et al. (2024) provides solid blueprints, building datasets using only attributes that would have been available at the decision date and carefully converting absolute dates to relative time spans. Such practices, while still not universal, are essential for ensuring a model's real-world validity. These data engineering steps are crucial for building a technically sound model; yet, a model that is inscrutable to its users has little practical value, which leads to the final component of the methodological pipeline: explainability.

### 4.2.4. Explainability (XAI): A Field in Transition

The final, and perhaps most user-facing, point of methodological divergence is explainability (XAI). Our findings show that while XAI is gaining traction, its application is inconsistent, creating an 'explainability gap' that separates transparent, trustworthy models from opaque black boxes. The most significant trend is the rise of SHAP (SHapley Additive exPlanations) as the new mainstream technique for both global and local model interpretation. It is increasingly valued for its game-theoretic foundation and model-agnosticism, with strong examples of its use in recent, high-quality studies (e.g., Razaghzadeh Bidgoli et al., 2024; Te, Liu, & Müller, 2023).

However, a large number of studies still rely on older, model-intrinsic importance metrics like Gini impurity or feature gain from tree-based ensembles (e.g., Gautam & Wattanapongsakorn, 2024; Abhinand & Poonam, 2022). While useful for a high-level overview, these methods can be biased and lack the crucial, per-prediction insight of SHAP.

Most concerning is the persistent explainability gap. We found that a substantial portion of the literature, particularly older or more technically-focused works, omits any formal feature-importance analysis, stopping at aggregate performance metrics and leaving stakeholders with an opaque "black box" prediction (e.g., Srivani et al., 2023; Belgaum et al., 2024). This limits trust and practical adoption. The field is clearly moving toward greater transparency, but treating XAI as an essential, rather than optional, component of the research process is a transition that is still underway.

## Table 5: Frequency and Application of AI/ML Algorithm Families Across the Corpus (Abridged, N=57)

This table presents a quantitative breakdown of the modeling techniques used across the 57-paper corpus, organized by algorithmic family. It reveals a clear hierarchy, with tree-based and boosting ensembles reigning as the dominant paradigms.

| Rank | Algorithm Family | Core Concept & Representative Algorithms | Prevalence (N=57) | Representative Studies (Citations) |
|---|---|---|---|---|
| 1 | Tree-based | Algorithms that create hierarchical, rule-based models. • Decision Tree (CART), Random Forest, Extra-Trees | 39 (72%) | Razaghzadeh Bidgoli et al. (2024), Takas et al. (2024), Font-Cot et al. (2025), Guerzoni et al. (2021) |
| 2 | Ensemble / Boosting | Combining multiple models to improve predictive performance. • Gradient Boosting, XGBoost, LightGBM, AdaBoost, CatBoost, Stacking | 37 (67%) | Setty et al. (2024), Samudra & Satya (2024), Gautam & Wattanapongsakorn (2024), Vasquez et al. (2023) |
| 3 | Linear / GLM | Models assuming a linear relationship between features and the target. • Logistic/Linear Regression, Ridge/Lasso, Elastic-net | 29 (54%) | Żbikowski & Antosiuk (2021), Dellermann et al. (2017), Liu et al. (2022), Schade & Schuhmacher (2023) |
| 4 | Kernel / Instance-based | Models making predictions based on feature similarity or distance. • Support Vector Machine (SVM), k-Nearest Neighbors (KNN) | 29 (53%) | Antretter et al. (2018), Cholil et al. (2024), Pasayat & Bhowmick (2021), Setty et al. (2024) |
| 5 | Neural Networks / Deep Learning | Biologically-inspired models with interconnected layers of nodes. • MLP/ANN, CNN, Bi-LSTM, TextCNN, LLM-agents | 19 (35%) | Chen. (2024), Ferrati et al. (2021), Srivani et al. (2023), Te, Liu, & Müller (2023) |
| 6 | Bayesian / Probabilistic | Models based on applying Bayes' theorem and probability distributions. • Naive Bayes, Bayesian Networks | 9 (16%) | Gujarathi et al. (2024), Al Rahma & Abrar-Ul-Haq (2024), Takas et al. (2024) |

*Note: This is an abridged version showing representative studies. For an exhaustive list of studies employing each algorithm family, please see the full table in Appendix C.*

## 4.3. Data as the Bedrock: Navigating the Trade-off Between Scale, Depth, and Fusion (RQ3)

The predictive power of any model is fundamentally constrained by the data it ingests. Our analysis of the data ecosystem (RQ3) reveals less a single point of convergence and more a landscape of competing strategic choices. We find the field is defined by a central trade-off between two divergent data strategies: a 'scale' strategy that prioritizes breadth using large venture databases, and a 'depth' strategy that prioritizes richness using bespoke data sources. The most sophisticated studies are now moving towards a third strategy—data fusion—to resolve this trade-off. However, we also find that all approaches are tempered by foundational challenges in data access, reproducibility, and ethics that limit the field's progress.

### 4.3.1. The Data Ecosystem: A Tale of Two Strategies (Scale vs. Depth)

Our analysis of 46 distinct raw data sources (see Table 6 for a summary and Appendix D for the full list) reveals a landscape defined by two divergent and competing philosophies of data collection.

The first, and most common, is the "scale" strategy, which prioritizes generalizability and large sample sizes. This approach relies on large, institutional Venture Databases, which are the primary data source in 28 of the 57 studies (49%). Crunchbase is the undisputed workhorse here, used for its massive scale (often millions of firms) and its structured, relational data on companies, funding, and exits (Gautam & Wattanapongsakorn, 2024; Kim et al., 2023). While this strategy allows for large-N, generalizable benchmarks, it often comes at the cost of feature depth, missing the nuanced, qualitative context behind a venture.

The second is the "depth" strategy, which crafts narrower but richer datasets to prioritize feature richness and capture signals invisible to institutional databases. This approach is evident in the use of more bespoke sources, such as:

- **Survey / Lab / Expert Panels (8 papers):** Used to source unique, theory-driven data like founder psychometric traits (Takas et al., 2024; Sabahi & Parast, 2020) or expert-scored evaluations from incubator programs (Gujarathi et al., 2024).
- **Unstructured Document Uploads (1 paper):** A pioneering study by Samudra & Satya (2024) demonstrates how LLMs can parse pitch-deck PDFs and financial spreadsheets to extract rich semantic features, emulating the qualitative due-diligence of a human analyst.
- **Registry / Filings (5 papers):** Studies focused on specific IP or financial data use sources like the USPTO patent database (Ross et al., 2021) or national company registries.

**Table 6: Taxonomy and Prevalence of Data Source Families in Startup Prediction (Abridged, N=57)**

This table provides a classification of the distinct raw data sources identified across the 57-study corpus, organized into nine families. It quantifies the field's bifurcated data strategy: a heavy reliance on large-scale Venture Databases for breadth, complemented by a long tail of bespoke sources.

| Data Source Family | Representative Raw Sources | Paper Count | Distinct Sources | Representative Studies (Citations) |
|---|---|---|---|---|
| Venture Databases | Crunchbase, AngelList, Seed-DB, IT Orange, Dealroom, Mattermark, CB Insights | 28 (49%) | 8 | Belgaum et al. (2024), Dellermann et al. (2017), Gautam & Wattanapongsakorn (2024), Ross et al. (2021), Setty et al. (2024) |
| Social-media / Open-web | Twitter, Facebook, LinkedIn, GitHub Stars, Telegram, Medium, Google Search hits, Alexa Web Traffic | 20 (35%) | 11 | Antretter et al. (2018), Razaghzadeh Bidgoli et al. (2024), Te, Liu, & Müller (2023), Żbikowski & Antosiuk (2021) |
| Survey / Lab / Expert Panel | Founder/Sustainability Surveys, Incubator/Expert Scores, Investor Crowd Ratings, Bahrain EO Survey | 8 (14%) | 6 | Takas et al. (2024), Font-Cot et al. (2025), Dellermann et al. (2017), Sabahi & Parast (2020) |
| Registry / Filings | USPTO PatentsView, SEC EDGAR Filings, EU VICO 2.0 Balance Sheets, National Company Registries | 5 (9%) | 4 | Ross et al. (2021), Ferrati et al. (2021), Liu et al. (2022) |
| Academic Benchmark Set | Kaggle "DPhi Startup", Kaggle "Startup Success", Public Crunchbase CSVs | 4 (7%) | 4 | Cholil et al. (2024), Gracy et al. (2024), Samudra & Satya (2024) |
| Hybrid Scrape / Composites | Venture Intelligence, StartupTalky, TechCrunch (News), Alexa Web-traffic | 4 (7%) | 4 | Abhinand & Poonam (2022), Belgaum et al. (2024), Razaghzadeh Bidgoli et al. (2024) |
| Token / Crowdfunding Platforms | Kickstarter, Wadiz, Tumblbug, Crowdy, ICOBench, ICOdrops | 2 (4%) | 6 | Dziubanovska et al. (2024), Eljil & Nait-Abdesselam (2024) |
| Search-API Enrichments | Google Custom-Search API | 1 (2%) | 1 | Samudra & Satya (2024) |
| Document Upload / Internal Files | Pitch-deck PDFs, Excel Financials | 1 (2%) | 2 | Samudra & Satya (2024) |

*Note: This is an abridged version showing representative studies. For an exhaustive list of studies employing each data source family, please see the full table in Appendix D.*

### 4.3.2. The Methodological Frontier: Data Fusion as the Synthesis Strategy

The most advanced research resolves the 'scale-versus-depth' dilemma through a powerful synthesis strategy: data fusion. Instead of choosing between breadth and richness, these studies programmatically combine multiple data types to create a superior, multi-modal information asset that captures the best of both worlds. Our analysis reveals that this is no longer a niche activity; it is the methodological frontier where the most significant performance gains are found. Table 7 provides a thematic typology of the studies in our corpus that employ a data fusion strategy.

The dominant fusion patterns involve enriching a core venture database with other sources to combine the best of both worlds—the sample size of "scale" with the feature richness of "depth." For example, Te et al. (2023) join Crunchbase firm data with scraped LinkedIn profiles to add a deep human-capital layer, while Ross et al. (2021) and Ferrati et al. (2021) perform cross-registry joins with USPTO patent records to inject an intellectual property dimension.

The consistent performance lift reported in these studies confirms a key finding: in startup prediction, richer data beats bigger data. The research consensus is that no single data source is sufficient, and competitive advantage lies in the thoughtful integration of disparate data types.

### Table 7: Thematic Typology of Data Fusion Strategies in Startup Prediction (N=31)

This table synthesizes the various data fusion patterns observed across the 31 studies in the corpus that combine data sources. It moves beyond a paper-by-paper listing to describe the core strategy, common data sources, and modeling value for each approach, illustrating the field's progression from simple data joins to sophisticated, multi-modal synthesis.

| Pattern Archetype | Description of Fusion Strategy | Core Rationale & Modeling Value | All Studies in Category |
|---|---|---|---|
| **1. Cross-Registry & Institutional Enrichment** | This foundational strategy involves enriching a primary venture database (typically Crunchbase) with other formal, structured data sources. The fusion is executed via programmatic joins on company names or unique IDs to link data from patent offices (USPTO), accelerator rosters (Seed-DB), or investor intelligence platforms (CB Insights). | This approach adds critical, non-financial dimensions like intellectual property depth and reputational signals from elite backers. It provides a more holistic firm profile than any single source allows, connecting funding data to tangible and intangible assets. | Abhinand & Poonam (2022), Antretter et al. (2018), Ferrati et al. (2021), Kim et al. (2023), Li, Y. et al. (2024), Misra, Jat, & Mishra (2023), Pasayat & Bhowmick (2021), Ross et al. (2021), Żbikowski & Antosiuk (2021) |
| **2. Web & Social Media Signal Integration** | This strategy augments structured firm data with dynamic, often unstructured, signals scraped from the open web and social media. It involves linking company profiles to their presence on platforms like LinkedIn and Twitter, or to news mentions and general web traffic. This requires significant pre-processing to handle unstructured text and time-series data. | Its purpose is to capture real-time market traction, public attention, and deep human-capital signals (e.g., founder experience from LinkedIn) that are absent from static databases. This provides a powerful, low-cost proxy for brand momentum and network effects. | Belgaum et al. (2024), Guerzoni et al. (2021), Razaghzadeh Bidgoli et al. (2024), Sharchilev et al. (2018), Te, Liu, & Müller (2023), Ungerer et al. (2021) |
| **3. NLP-Driven Document & Textual Fusion** | This advanced strategy uses Natural Language Processing (NLP) and Large Language Models (LLMs) to extract structured features, sentiment scores, or semantic vectors from unstructured text. Primary sources include company descriptions, news articles, and, most recently, user-uploaded pitch decks. | This fusion unlocks the rich, qualitative information "locked" inside documents. It allows the model to incorporate nuanced signals such as emotional tone in a pitch, specific ESG narratives in a campaign, or key business plan concepts, thereby emulating the qualitative assessment of a human analyst. | Goossens et al. (2023), Samudra & Satya (2024), William et al. (2022), Zhang & Lau (2023) |
| **4. Human-in-the-Loop & Primary Data Composites** | This archetype involves creating bespoke datasets by combining primary data sources that require direct human input. It fuses expert-scored survey data from incubator panels, founder psychometric test results, crowdsourced ratings, and semi-structured interview data. The fusion is often complex, sometimes occurring | This approach is essential for testing specific, theory-driven hypotheses (e.g., the impact of founder personality) and capturing subjective, "soft" signals that are impossible to scrape. It yields deep, context-rich datasets that often lead to very high performance in niche settings, | Al Rahma & Abrar-Ul-Haq (2024), Dellermann et al. (2017), Font-Cot et al. (2025), Gujarathi et al. (2024), Sabahi & Parast (2020), Takas et al. (2024) |

probabilistically within the model itself (e.g., a Bayesian Network).    albeit with limited generalizability.

Note: This table is exhaustive, listing all 31 studies from the corpus that explicitly combine more than one distinct raw data source. Each study is classified under the archetype that best represents its primary fusion contribution.

### 4.3.3. Foundational Challenges: Data Structure, Access, and Ethics

While data fusion offers a path forward, our analysis reveals three foundational challenges that underpin all data strategies and define the field's practical limitations. These issues of data structure, access, and ethics, observed consistently across our corpus, represent significant hurdles to building a truly robust, reproducible, and responsible body of research.

First, data structure is evolving in complexity. While early work often relied on simple tabular files, the norm in more recent and sophisticated studies involves complex relational joins, the integration of unstructured text with structured numerics, and the construction of longitudinal panels to capture temporal dynamics (e.g., Ferrati et al., 2021; Srivani et al., 2023; Te et al., 2023). This increasing complexity places higher demands on researchers' data engineering capabilities.

Second, data access is highly fragmented, presenting a significant hurdle to reproducibility. Our analysis confirms that sources range from fully Open datasets (e.g., Kaggle datasets used by Cholil et al., 2024), to Freemium services requiring academic licenses (e.g., Crunchbase, the most common source), to entirely Proprietary data from accelerators or private surveys (e.g., Takas et al., 2024). This "access fragmentation" means that while modeling code is increasingly shared on platforms like GitHub, the underlying raw data often remains inaccessible, severely limiting direct replication and verification of findings.

Finally, ethics and privacy are nascent but critical concerns. The majority of studies in our corpus using publicly available firm-level data make no mention of formal IRB review. While studies involving primary data from human subjects are beginning to document informed consent (e.g., Takas et al., 2024), a significant gap remains in the discussion of PII (Personally Identifiable Information) and GDPR compliance, especially in studies that scrape founder information from social media (e.g., Te et al., 2023). These foundational challenges of data access, reproducibility, and ethics not only limit current research but also define the most urgent priorities for building a more robust and responsible field, a point we will return to in our discussion of future research agendas.

# 5. A Conceptual Framework for Principled Startup Success Forecasting

The preceding 'Results' section systematically deconstructed the landscape of AI-based startup prediction, revealing a field defined by a central paradox: a convergence on common tools but a stark divergence in methodological rigor. We have demonstrated how this ad-hoc approach—characterized by fragmented success definitions, atheoretical features, and inconsistent validation—limits the comparability, reliability, and practical utility of existing work. To address these identified gaps directly, we now move from deconstruction to prescription. We propose the Systematic AI-driven Startup Evaluation (SAISE) Framework, a novel, five-stage conceptual roadmap designed to guide future research from ad-hoc prediction toward principled, reproducible, and impactful evaluation.

## 5.1. The Systematic AI-driven Startup Evaluation (SAISE) Framework

We introduce the **Systematic AI-driven Startup Evaluation (SAISE) Framework** as a prescriptive roadmap for researchers. The name is chosen with intent, with each component designed to counter a specific weakness identified in our review:

- **Systematic:** because it directly counters the ad-hoc approaches prevalent in the literature, replacing them with a structured, five-stage pipeline that guides the researcher from problem formulation to actionable interpretation.
- **AI-driven:** because it centers modern AI techniques as the core engine for prediction, but mandates the best practices required to harness their power effectively, thus closing the gap between common and best practice in modeling.
- **Evaluation:** because its goal extends beyond simple classification accuracy to a more holistic assessment, directly addressing the field's explainability gap and the need for risk-aware, decision-aligned analytics.

The SAISE framework is built on four core principles. Each principle is derived directly from a foundational gap we identified in our review, ensuring that the framework is not just a proposal but an evidence-based solution.

1. **Stage-Awareness:** To resolve the Definitional Gap (Section 4.1.3), the framework mandates that the predictive objective be explicitly tailored to a startup's developmental stage (e.g., pre-seed, growth). This directly addresses the "pre-seed prediction problem" and forces a clear, context-appropriate definition of "success" from the outset.
2. **Data-Centric Synthesis:** To address the Data Ecosystem Gap (Section 4.3), this principle moves beyond single-source reliance. It formalizes data fusion as a core strategic process, prescribing the enrichment of large-scale 'scale' data with deep, contextual 'depth' signals to create superior, multi-modal information assets.
3. **Principled Feature Engineering:** To bridge the Theoretical Gap (Section 4.1.1), this principle advocates for a theory-informed approach to feature creation. It mandates the development of dynamic, relational, and temporally-aware features grounded in established entrepreneurship theory, moving beyond the use of atheoretical, convenience-based variables.

4. **End-to-End Methodological Rigor:** To close the Methodological Gap (Section 4.2), this principle elevates the field's best practices from rare exceptions to mandatory standards. It requires robust validation (e.g., nested CV), systematic hyper-parameter tuning, and modern XAI as non-negotiable components of the research pipeline.

### 5.2. The Five Stages of the SAISE Framework

The following sections will detail each of the five stages of this framework. As visually represented in Figure 2, the five stages are organized into three distinct conceptual phases: a Pre-processing/Setup phase for problem and data formulation (Stage 1), a core Model Development/Processing phase for the technical pipeline (Stages 2-4), and a final Post-processing/Interpretation phase focused on insight generation (Stage 5). This structured pipeline is designed to guide researchers from initial conception to actionable conclusion, providing a clear and actionable guide for future research. By adhering to this sequence, researchers can systematically avoid the ad-hoc pitfalls that currently fragment the field.

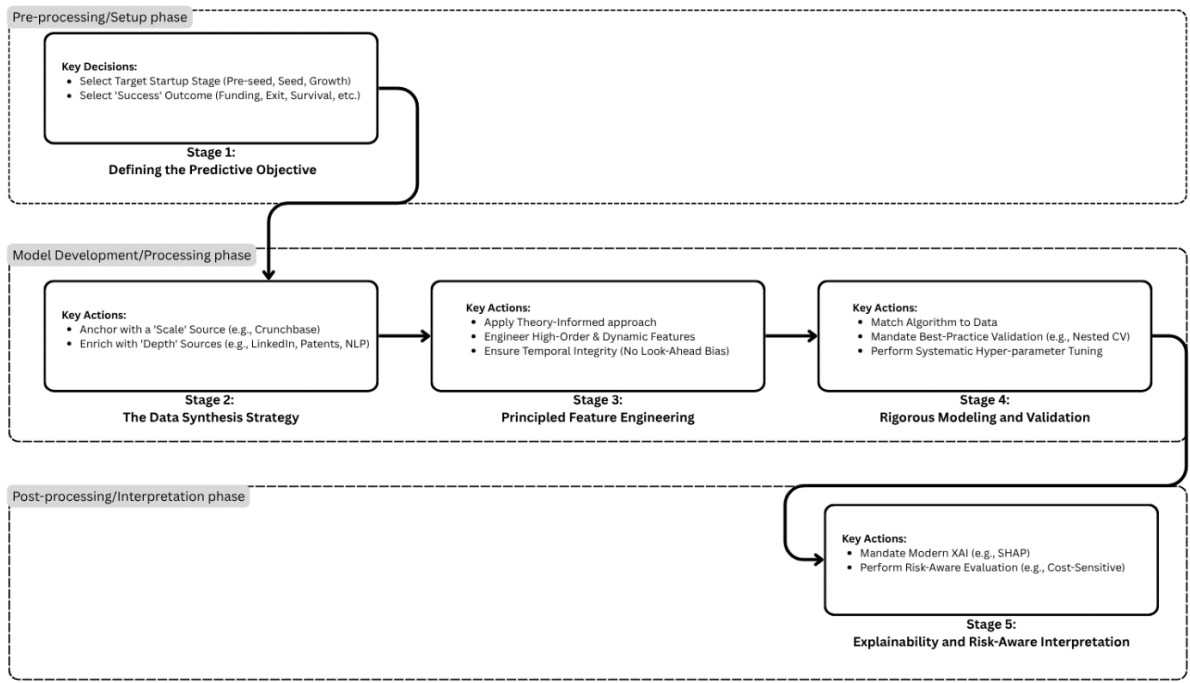

*Figure 2. The Systematic AI-driven Startup Evaluation (SAISE) Framework. This visual roadmap outlines the five sequential stages designed to guide principled research in startup success forecasting. The stages are grouped into three distinct phases to emphasize a structured workflow, moving from problem formulation (Stage 1), through technical execution (Stages 2-4), to final insight generation (Stage 5).*

### 5.2.1. Stage 1: Defining the Predictive Objective

The foundational stage of the SAISE framework addresses the most critical source of heterogeneity identified in our review: the lack of a standardized problem definition. As established in our analysis of target variables (Section 4.1.3), a model's purpose, architecture, and practical utility are inextricably linked to the specific context and outcome it is designed to predict. This initial stage, therefore, compels researchers to make two explicit, interdependent decisions before any data is sourced or features are engineered, thereby preventing common category errors and ensuring the research question is well-posed.

The first decision is the selection of the target startup stage. Our findings reveal a significant "pre-seed prediction problem," where models developed on data from later-stage ventures are implicitly inapplicable to the earliest stages (Takas et al., 2024; Razaghzadeh Bidgoli et al., 2024). The SAISE framework mandates that researchers first define their temporal focus, choosing from distinct phases such as:

- **Pre-Seed / Idea Stage:** Ventures with no formal funding history and often no product. Prediction at this stage must rely on non-financial signals like founder attributes, idea quality, and psychometric data.
- **Seed / Early-Stage:** Ventures that have secured initial funding but have limited operational or financial history. Prediction can begin to incorporate early funding signals, team growth, and initial product traction.
- **Growth / Scale-up Stage:** Established ventures with significant funding, operational history, and market presence. Prediction can leverage rich time-series data on financial performance, market expansion, and industry dynamics.

The second, and equally critical, decision is the selection of the "success" outcome. Our analysis in Section 4.1.3 and Table 4 demonstrated a fragmented landscape of target variables, from funding milestones to long-term survival. This choice must be logically consistent with the selected startup stage. For example, modeling an "IPO Exit" (Kim et al., 2023) is a valid objective for growth-stage firms but is nonsensical for pre-seed ventures. The framework requires researchers to select and clearly justify their target from the primary themes we identified:

- **Funding-Milestone Achievement** (e.g., securing a Series-A round; Te et al., 2023).
- **Exit Event / Firm Status** (e.g., IPO/M&A vs. closure; Setty et al., 2024; Ross et al., 2021).
- **Survival / Sustainability** (e.g., 5-year survival probability; Font-Cot et al., 2025; Takas et al., 2024).
- **Valuation / Investor Return** (e.g., market-cap regression or unicorn classification; Chen., 2024; Abhinand & Poonam, 2022).
- **Financial Distress Avoidance** (e.g., predicting bankruptcy; Liu et al., 2022).

By forcing this initial two-part declaration, Stage 1 establishes a disciplined foundation for the entire research process. It ensures that the subsequent choices regarding data, features, and algorithms are coherently aligned with a clearly articulated and contextually appropriate research objective, directly addressing the definitional ambiguity that currently limits the comparability and synthesis of findings across the field.

### 5.2.2. Stage 2: The Data Synthesis Strategy

With a clearly defined objective from Stage 1, the next step is to construct the data asset. The SAISE framework moves beyond the simple inventorying of sources to advocate for a deliberate data synthesis strategy. Our analysis in Section 4.3 revealed that research in this domain is bifurcated into two approaches: a "scale" strategy relying on large but potentially shallow institutional databases, and a "depth" strategy using narrower, bespoke datasets to capture unique signals. We found that the most advanced and performant studies resolve this trade-off through data fusion, a process we elevate from a niche activity to a core component of the framework. Stage 2 provides a blueprint for this synthesis.

The strategy begins with anchoring the study in a 'scale' source. The undisputed workhorse of the field is the large-scale venture database, with Crunchbase being the most prominent example. These platforms provide the necessary sample size (N) for building generalizable benchmark models and offer structured, relational data on core entities like funding rounds, company profiles, and exit events (Gautam & Wattanapongsakorn, 2024; Kim et al., 2023). However, as our findings suggest, relying on this data alone often misses the nuanced, qualitative context that drives success.

Therefore, the second and crucial step is to systematically enrich the scale source with one or more 'depth' sources. The goal is to create a proprietary, multi-modal information asset that combines the statistical power of large-N data with the predictive lift of high-fidelity signals. Based on the successful fusion patterns identified in our review (Table 7), the framework recommends several key enrichment pathways:

- **Human Capital Depth:** Augmenting firm-level data with founder-level data scraped from professional networks like LinkedIn. This injects granular information on founder experience, skills, and network connections, a strategy proven effective by Te et al. (2023).
- **Intellectual Property Depth:** Cross-referencing company data with patent registries like the USPTO. This allows for the creation of sophisticated, time-series features related to a firm's technological innovation and defensibility (Ferrati et al., 2021; Ross et al., 2021).
- **Qualitative & Semantic Depth:** Employing Natural Language Processing (NLP) to parse unstructured documents like pitch decks or news articles. This unlocks rich, qualitative information—such as strategic framing or sentiment—that is invisible to structured databases, a frontier explored by Samudra & Satya (2024).
- **Theory-Driven Depth:** Integrating primary data from surveys or expert panels designed to measure specific theoretical constructs (e.g., founder psychometrics or team dynamics), as demonstrated by Takas et al. (2024) and Font-Cot et al. (2025).

### 5.2.3. Stage 3: Principled Feature Engineering

Following the synthesis of a multi-modal data asset in Stage 2, Stage 3 focuses on the critical task of feature engineering: the process of transforming raw data into predictive signals. This stage is arguably where the most significant competitive advantage is created. The SAISE framework advocates for a disciplined and creative approach that moves beyond simply using variables as they appear in a source file. It is designed to resolve the methodological tension between theory-driven and data-driven approaches (Section 4.1.1) and to operationalize the key insight that dynamic, contextual features consistently outperform static ones (Section 4.1.2).

First, the framework prescribes a 'theory-informed' approach to feature creation. This approach resolves the divide between studies that are rigorously grounded in theory but lack scale (e.g., Font-Cot et al., 2025) and those that are large-scale but atheoretical (e.g., Cholil et al., 2024). Instead of starting with theory in a vacuum, researchers should use established management and entrepreneurship theories—such as the Resource-Based View (RBV), Network Theory, or Upper Echelons Theory—as a conceptual lens to guide the extraction and construction of features from their large-scale, fused dataset. For example, a researcher could use Network Theory to guide the creation of centrality metrics from scraped LinkedIn data, as done by Kim et al. (2023) at the industry level and Li et al. (2024) at the cohort level.

Second, this stage mandates the engineering of high-order, dynamic features. Our analysis consistently showed that static attributes (e.g., total funding) are often less predictive than dynamic indicators of momentum and context. Researchers should therefore prioritize the creation of features that capture:

- **Relational and Interaction Dynamics:** These features model the interplay between different entities, moving beyond simple firm-level attributes. Examples include engineering cohort-level network features that capture the relational dynamics between startups in an accelerator batch (Li et al., 2024).
- **Momentum and Velocity:** These features capture the rate of change rather than a static state. Examples include Funding Velocity (the time between rounds), the growth rate of social media followers, or the rate of patent filings over time (Ferrati et al., 2021).
- **Contextual Embeddings:** These features place a startup within its ecosystem. Examples include calculating an Industry Convergence Level (Kim et al., 2023) or a firm's centrality within its accelerator cohort (Li et al., 2024).

Finally, a non-negotiable principle of this stage is ensuring temporal integrity by preventing look-ahead bias. This is a crucial safeguard for creating models with real-world validity. As demonstrated by the methodologically rigorous approaches of Li et al. (2024) and Żbikowski & Antosiuk (2021), all features must be constructed using only information that would have been available at the point of prediction. This involves, for instance, avoiding post-decision funding data and carefully converting absolute dates (e.g., "2023-05-10") into relative time spans (e.g., "age in days at prediction").

### 5.2.4. Stage 4: Rigorous Modeling and Validation

After engineering a high-quality feature set, the focus shifts to the core of the predictive task: the selection, training, and validation of the machine learning model. This stage is paramount for producing trustworthy and reproducible results, yet our review in Section 4.2 revealed it as the site of the most significant "Gap Between Common Practice and Best Practice." Many studies fall short of established standards for validation and tuning, undermining the credibility of their findings. The SAISE framework, therefore, elevates these best practices from rare exceptions to mandatory protocols.

First, the choice of algorithm must be a deliberate, justified decision matched to the data's structure and the research objective, not merely a default. Our analysis confirmed the "unquestioned reign of tree-based ensembles" like XGBoost, LightGBM, and Random Forest, which remain the optimal choice for the structured, tabular data that characterizes much of this field (Cholil et al., 2024; Gautam & Wattanapongsakorn, 2024). Their proven ability to handle non-linear relationships and provide intrinsic feature importance makes them a powerful default. However, as research moves toward the multi-modal data synthesis advocated in Stage 2, the algorithmic toolkit must expand. For datasets fusing tabular data with unstructured text or images, more sophisticated architectures like hybrid or two-tower neural networks are required to process heterogeneous data streams effectively (Chen., 2024). Furthermore, looking to the research frontier, emerging AI Agentic frameworks, which combine the reasoning capabilities of Large Language Models (LLMs) with traditional predictive tools, present a new and promising path for integrating complex, qualitative assessments directly into the modeling pipeline.

Second, and most critically, the framework mandates best-practice validation protocols to ensure model generalization and prevent optimistic bias. A simple train/test hold-out split, while common (e.g., Zhang & Lau, 2023), is highly susceptible to sampling bias and is insufficient for rigorous academic work. Instead, we prescribe a hierarchy of more robust techniques:

- **k-Fold Cross-Validation (CV):** This should be the minimum standard for performance estimation. By systematically partitioning the data into multiple train-and-test folds, it provides a much more stable and reliable estimate of a model's performance on unseen data than a single split (Gangwani et al., 2023).
- **Nested Cross-Validation:** For any study involving hyper-parameter tuning, a properly nested CV scheme is non-negotiable. The "outer loop" splits the data for performance evaluation, while a completely separate "inner loop" is used on the training portion of each outer fold to find the best hyper-parameters. This strict separation, though computationally expensive and rarely implemented correctly, is the gold standard for preventing information from the tuning process "leaking" into the final performance estimate. The work of Te et al. (2023) stands as a key exemplar of this best practice.
- **Temporal Validation:** For any objective involving a time-series or forecasting element (e.g., predicting survival or a future funding round), a strict temporal split or rolling-origin validation is essential. This ensures the model is only trained on past data to predict future events, simulating a real-world deployment scenario and rigorously guarding against look-ahead bias (Wang et al., 2022).

Finally, systematic and transparent hyper-parameter tuning must be treated as a core scientific step. Our review found that an alarming 50% of studies either rely on default library parameters or perform minimal manual adjustments (e.g., Kuznietsova & Grushko, 2019; Abhinand & Poonam, 2022). This "minimalist mindset" makes robust model comparisons impossible and severely limits performance. The SAISE framework requires the use of formal, automated search strategies—such as exhaustive Grid Search, efficient Random Search, or intelligent Bayesian Optimization—coupled with a clear report of the search space and the optimal parameters found.

### 5.2.5. Stage 5: Explainability and Risk-Aware Interpretation

The final stage of the SAISE framework ensures that a technically sound model becomes a practically valuable one. A prediction, no matter how accurate, is of little use if it is not understood, trusted, and contextualized for real-world decisions. This stage, therefore, moves beyond reporting aggregate performance metrics to focus on model interpretation and risk-aware evaluation, directly addressing the "explainability gap" and the need for decision-aligned analytics identified in our review (Sections 4.2.4 and 4.1.3).

First, the framework mandates the use of modern Explainable AI (XAI) techniques. Our analysis revealed that a substantial portion of the literature, particularly older or more technically-focused works, stops at reporting accuracy scores, leaving stakeholders with an opaque "black box" (e.g., Srivani et al., 2023). To close this gap, we prescribe the adoption of SHAP (SHapley Additive exPlanations) as the new mainstream standard. Valued for its game-theoretic foundation and model-agnosticism, SHAP provides several critical layers of insight unavailable in older methods like Gini impurity:

- **Global Importance:** It offers an unbiased view of which features are driving model predictions overall.
- **Local Interpretation:** Crucially, it can explain individual predictions, showing which features contributed positively or negatively to the outcome for a specific startup. This is essential for building user trust and generating actionable insights (Wang et al., 2022; Te et al., 2023).

Second, the framework calls for a risk-aware interpretation of model output. Standard classification metrics like accuracy can be misleading, especially in investment contexts where the cost of different errors is highly asymmetric. An investor fears a False Positive (investing in a startup that fails) far more than a False Negative (passing on a startup that succeeds). The framework therefore calls for this reality to be reflected in two complementary ways: first, through the careful selection of evaluation metrics, and second, through the direct adoption of cost-sensitive modeling techniques.

1. **Evaluation Metrics:** Researchers should prioritize and report on metrics that are sensitive to class imbalance and error costs, such as **Precision**, **Recall**, and the **F1-score**, rather than relying solely on overall accuracy.
2. **Cost-Sensitive Modeling:** For the most advanced applications, the learning objective itself can be redefined to align with financial risk. The cost-sensitive relabeling approach introduced by Setty et al. (2024), which uses the MetaCost algorithm, serves as a powerful exemplar. By training the model to explicitly minimize expected financial loss rather than classification error, this technique fundamentally shifts the goal from building an accurate predictor to building a risk-optimized decision-support tool.

# 6. Discussion

Our systematic review was motivated by a central question: In an era of powerful AI, what constitutes best practice for the evaluation of startups? The findings presented in the preceding sections reveal a vibrant and rapidly maturing research domain, yet one marked by significant methodological inconsistencies. The field is at a critical inflection point, moving from early, exploratory models to more sophisticated and potentially impactful decision-support tools. This discussion synthesizes our findings to first, diagnose the current state of the field; second, articulate the theoretical and practical implications; and third, propose a concrete research agenda to guide the field toward a more rigorous and coherent future.

### 6.1. Synthesis of Key Findings: A Field of Converging Practice and Diverging Rigor

Our analysis paints a clear picture of the current paradigm in AI-driven startup evaluation. At its core, the field is defined by three key characteristics: a remarkable convergence on a common set of tools and data, a persistent gap between common practice and methodological best practice, and an emerging frontier driven by synthesis and context.

First, there is an undeniable convergence on a dominant methodological toolkit. The typical study in this domain leverages a large-scale venture database, primarily Crunchbase, as its source of "scale." It populates its models with a core triad of feature families—Funding History, Team/Founder Attributes, and Market/Sector Tags—and employs tree-based ensemble algorithms like XGBoost or Random Forest as the predictive engine. This de facto "standard model" provides a valuable and generalizable baseline, but its ubiquity also highlights the field's heavy reliance on a narrow set of readily available, structured data sources and off-the-shelf algorithms.

Second, despite this convergence, our findings reveal a significant divergence in methodological rigor. We identified a persistent "gap between common practice and best practice" across the entire modeling pipeline. This is most evident in the frequent lack of theory-informed feature engineering, the rarity of robust validation techniques like nested cross-validation, and the often-overlooked "explainability gap" where models are presented as black boxes. This divergence means that while many studies appear similar on the surface, their true reliability and scientific validity differ immensely.

Finally, our review illuminates the methodological frontier where the most impactful research is occurring. This frontier is defined by a single concept: synthesis. The most sophisticated studies are moving beyond the standard model by actively fusing disparate data sources (e.g., combining Crunchbase with patent data as in Ross et al., 2021), integrating management theory with data-driven feature engineering, and combining raw predictive accuracy with deep, risk-aware explainability.

This approach is more than just a technical exercise; it represents a conceptual shift. It embodies the principle of "human-AI symbiosis" (Jarrahi, 2018), where the analytical power of the machine to process vast "hard" datasets is combined with human-centric "soft" knowledge drawn from theory and contextual data. Furthermore, it moves the field toward a model of "machines augmenting entrepreneurs" (Shepherd & Majchrzak, 2022), where the goal is not to replace human judgment but to enhance it with richer, more nuanced insights. These pioneering efforts, exemplified by the risk-aware models of Setty et al. (2024), confirm a key insight: competitive advantage is no longer found in simply applying an algorithm, but in the thoughtful synthesis of diverse data and methods to create a more holistic, context-aware, and trustworthy assessment of a venture.

### 6.2. Gaps and Limitations in the Current Literature (RQ4)

While the field is advancing rapidly, our systematic review identified four significant and recurring gaps that limit the reliability, comparability, and practical utility of current research. These gaps, which span the entire research process from problem formulation to data sourcing, provide the core motivation for the principled, systematic approach advocated by our SAISE framework.

**1. The Definitional Gap: A Fragmented Understanding of "Success"**
The most fundamental weakness is the absence of a standardized definition for the primary outcome variable: "startup success." As detailed in Section 4.1.3 and Table 4, success is variously defined as achieving a funding milestone, executing an exit, long-term survival, or avoiding financial distress. This fragmentation means that a model optimized to predict "Series-A funding" (e.g., Te et al., 2023) is being trained on a fundamentally different task than one designed to predict "10-year survival" (e.g., Ferrati et al., 2021). This lack of a consistent ground truth severely limits the ability to compare model performance across studies and to synthesize findings into a coherent body of knowledge.

**2. The Theoretical Gap: The Primacy of Convenience over Principled Feature Engineering**
A critical limitation, highlighted in Section 4.1.1, is the prevalent "theory-practice divide" in how predictor variables are selected and engineered. The majority of studies adopt a convenience-based, "data-driven" approach, utilizing features simply because they are available in a given dataset (e.g., Cholil et al., 2024). While pragmatic, this often leads to atheoretical models that may capture statistical correlations without providing deeper explanatory insight. In contrast, the few studies that are explicitly "theory-informed" (e.g., Font-Cot et al., 2025) often rely on small-scale, primary data collection. As argued by Giuggioli & Pellegrini (2022), AI should be seen as an enabler that connects theory with practice, not as a tool that widens the gap. There is a clear and urgent need to bridge this chasm by using established entrepreneurship theories (e.g., RBV, Network Theory) to guide the engineering of features from large-scale, secondary data sources, as prescribed in Stage 3 of our framework.

### 3. The Methodological Gap: A Chasm Between Common Practice and Best Practice

Our analysis in Section 4.2 revealed a troubling and persistent gap between common methodological choices and established best practices for building trustworthy AI models. Key weaknesses include:

- **Superficial Validation:** The over-reliance on simple train/test splits, which are prone to sampling bias, instead of more robust k-fold or nested cross-validation protocols.
- **Negligent Hyper-parameter Tuning:** An alarming number of studies rely on default algorithm parameters, severely limiting model performance and rendering comparisons invalid (e.g., Kuznietsova & Grushko, 2019; Abhinand & Poonam, 2022).
- **Pervasive Look-Ahead Bias:** Many studies fail to implement strict temporal validation, using features that would not have been available at the time of prediction, thus producing overly optimistic results with no real-world validity.
- **The Explainability Gap:** A substantial portion of the literature treats models as "black boxes," failing to provide feature-importance or local-level explanations (e.g., using SHAP), which limits trust and practical adoption.

### 4. The Data Ecosystem Gap: Over-Reliance, Access, and Ethics

Finally, as detailed in Section 4.3, the field faces foundational challenges related to its data ecosystem. There is a clear over-reliance on a single data source family—Venture Databases, and specifically Crunchbase—which risks creating a monoculture of research questions and features. This is compounded by issues of data access and reproducibility; while code is often shared, the underlying proprietary or scraped data is not, making direct replication impossible. Furthermore, nascent but critical ethical concerns regarding the use of personally identifiable information (PII) scraped from social media and the general lack of formal IRB review for studies using publicly available but sensitive data remain largely unaddressed.

### *6.3. Implications of the Findings*

The gaps and frontiers identified in this review carry significant implications for the primary stakeholders in the AI-driven startup evaluation ecosystem. Our findings offer a roadmap not only for advancing the academic state-of-the-art but also for improving the practical application of these powerful technologies.

### 1. Implications for Researchers

This review serves as a clear call to action for the academic community to elevate the standards of research in this domain. The path forward is not simply to build more models, but to build them more thoughtfully. Researchers should:

- **Advance from Convenience to Rigor in Methods and Features:** The field's reliance on readily available, static features from single data sources and the use of superficial validation techniques represent significant limitations. The "gap between common practice and best practice" must be closed. Future work should prioritize the engineering of dynamic, relational, and contextual features derived from the fusion of multiple data sources, as these have been shown to provide superior predictive power.

Simultaneously, adopting the principles outlined in our SAISE framework—including robust validation with nested CV, systematic tuning, and modern XAI—should become the new standard for publication.

- **Tackle the "Pre-Seed Prediction Problem":** There is a clear disconnect between the needs of practitioners (evaluation of pre-seed ventures) and the capabilities of most current models. This represents a major opportunity for high-impact research focused on developing novel, non-financial signals for the earliest stages of a venture's life. This could involve integrating founder psychometrics (Takas et al., 2024) or leveraging "Hybrid Intelligence" to combine machine analysis with human intuition for assessing "soft" signals that exist before the first financial data is generated (Dellermann et al., 2017).
- **Integrate Theory with Data Science:** Researchers must actively work to bridge the theory-practice divide. This involves using established entrepreneurship and management theories not just as post-hoc justifications, but as a guiding lens for hypothesis formulation, data synthesis, and feature engineering.

## 2. Implications for Practitioners (Investors, VCs, and Accelerators)

For practitioners, our findings suggest that AI/ML is a powerful but not magical tool. They should become critical and informed consumers of these technologies.

- **Question the "Ground Truth":** When evaluating an AI tool, the most important question to ask is: "How was 'success' defined?" A model trained to predict Series-A funding is a tool for identifying fundraising potential, not necessarily long-term business viability. Practitioners must ensure the model's objective aligns with their own investment thesis.
- **Demand Explainability, Not Just Accuracy:** A black-box prediction with a high accuracy score is of limited use. Practitioners should demand models that offer transparent, per-prediction explanations (e.g., via SHAP), allowing them to understand the "why" behind a recommendation and integrate it with their own domain expertise.
- **Recognize the Value of Proprietary Data:** The finding that "richer data beats bigger data" implies that organizations with access to unique, proprietary data (e.g., internal deal flow notes, expert evaluation scores, detailed founder interviews) are best positioned to build a true competitive advantage through AI. The fusion of this deep, internal data with large-scale external data is the most promising path forward.

## 3. Implications for Entrepreneurs

For founders and their teams, this review provides a data-backed mirror reflecting what the AI-driven investment ecosystem values and measures.

- **The Signals That Matter:** The features consistently identified as most predictive—such as founder's prior experience, team completeness, clear funding cadence, and strong digital traction—offer a clear roadmap of the signals that need to be actively managed and communicated to investors.
- **Narrative and Data Are Intertwined:** The rise of NLP-based analysis means that the narrative of a pitch deck and the company's online presence are no longer just "soft" assets; they are machine-readable data points that directly feed into evaluation models. Crafting a coherent and compelling story is a data-driven imperative.

### 6.4. Limitations of this Review

While this systematic literature review was conducted with a rigorous and comprehensive protocol, it is essential to acknowledge its inherent limitations. These boundaries define the scope of our conclusions and provide context for the interpretation of our findings.

First, the **scope of our search**, though extensive, is not exhaustive. Our search was confined to peer-reviewed journal articles and conference papers from specific academic databases (Web of Science and Scopus). Consequently, we may have missed relevant contributions from adjacent fields not well-covered by our search string (e.g., pure finance, computational social science), as well as valuable insights from grey literature such as books, dissertations, or influential industry reports. Our use of English-language papers only also introduces a potential geographic and linguistic bias.

Second, our findings are susceptible to publication bias, a common challenge in all literature reviews. Academic publishing tends to favor studies with statistically significant or novel positive results. It is therefore likely that the literature we reviewed over-represents successful modeling attempts, while studies that found certain features or algorithms to be ineffective are under-represented. The true state of the field may include more null results and failed experiments than is visible in the published record.

Third, while we employed a systematic protocol for data extraction and analysis, the process of thematic synthesis inherently involves a degree of interpretive judgment. The categorization of features, algorithms, or data sources into the conceptual families presented in our Results section was guided by our deep reading of the corpus, but other researchers might have reasonably grouped them differently.

Finally, this review reflects a snapshot of a rapidly evolving field. The pace of innovation in both AI/ML and the startup ecosystem is extraordinary. New techniques, particularly those involving Large Language Models and AI agents, are emerging so quickly that any review is, by definition, a look at the recent past. The landscape is likely to have shifted even by the time of this publication. These limitations, however, do not diminish the core findings regarding the persistent methodological gaps and the need for a principled framework, which we believe are foundational and will remain relevant as the field continues to mature.

### 6.5. A Future Research Agenda

Based on the findings and limitations identified in this review, we propose a concrete agenda for future research designed to push the field beyond its current frontiers. This agenda moves from addressing immediate gaps to exploring next-generation methodologies, aiming to build AI systems for startup evaluation that are more robust, insightful, and aligned with real-world complexities.

**1. Tackling the Pre-Seed Prediction Problem**
The most significant practical gap is the field's inability to reliably evaluate pre-seed ventures.

Future research should pivot away from models dependent on financial history and focus exclusively on this early stage. This requires developing and validating novel, non-financial features by:

- **Advancing NLP on Idea-Stage Texts:** Using advanced NLP and LLMs to extract nuanced signals of idea quality, novelty, and market positioning directly from unstructured business plans or idea descriptions.
- **Integrating Founder Psychometrics:** Expanding on the work of Takas et al. (2024) by systematically collecting and incorporating psychometric data (e.g., resilience, cognitive agility, risk propensity) to model the "jockey" when the "horse" is still just an idea.

**2. Bridging the Chasm**: A Systematic Review of Theories for Feature Construction
A key weakness we identified was the theory-data divide. To address this systematically, we call for a dedicated systematic literature review focused on entrepreneurial and management theories. The goal of such a review would be to catalog theoretical constructs known to influence venture success (e.g., from RBV, effectuation theory, network theory) and map them to potentially accessible data sources. This would produce an invaluable, theory-backed feature framework, providing a common reference point for researchers and helping to solve the problem of atheoretical, convenience-based feature selection.

**3. From Prediction to Causal and Counterfactual Inference**
The current paradigm is overwhelmingly predictive ("Will this startup succeed?"). The next leap in value will come from models that can provide causal and counterfactual insights ("Why will this startup succeed, and what could be changed to increase its chances?"). We encourage research that incorporates techniques from the causal inference literature to move towards answering questions like:

- What is the causal impact of having a founder with prior exit experience?
- What would be the predicted outcome if a startup had chosen a different market entry strategy?

**4. The Next Frontier: Agentic AI for Holistic Startup Evaluation**
Finally, the emergence of powerful Large Language Models opens up an entirely new research frontier centered on AI Agentic frameworks. These systems, which can combine logical reasoning, tool use, and access to vast knowledge bases, have the potential to automate and enhance the entire SAISE framework. We envision a future research agenda focused on building and testing agents that can:

- **Automate Data Synthesis:** An agent could be tasked to "find all relevant data about Startup X," automatically scraping social media, parsing news articles, and querying patent databases to build a rich, multi-modal profile on the fly.
- **Fuse Qualitative Theory with Quantitative Data:** An agent could be given a knowledge base of entrepreneurial theories (as identified in Agenda Point #2) and a quantitative dataset. It could then be prompted to "Evaluate this startup according to the principles of the Resource-Based View," providing a reasoned, qualitative assessment that complements the purely quantitative prediction.

Pursuing these research directions will not only address the limitations of the current literature but will also accelerate the development of the next generation of AI-driven tools—tools that are not just predictors, but genuine partners in the complex art and science of startup evaluation.

# 7. Conclusion

The integration of Artificial Intelligence into the high-stakes world of venture capital and startup evaluation represents a significant technological shift, yet the academic research underpinning this transition has remained methodologically fragmented. The ad-hoc application of models, often driven by data convenience rather than principled strategy, has limited the comparability, reliability, and practical utility of much of the existing work. This systematic literature review was conducted to deconstruct this landscape, identify critical gaps, and synthesize a clear path forward for researchers.

Our review revealed a field defined by a central paradox: a strong convergence on a common set of tools—namely, venture databases and tree-based ensembles—but a stark divergence in methodological rigor. We identified significant weaknesses in how success is defined, how features are engineered, and how models are validated and interpreted. In response, this paper's primary contribution is the proposal of the Systematic AI-driven Startup Evaluation (SAISE) Framework. This five-stage, sequential pipeline provides a prescriptive roadmap designed to directly address the identified gaps. By mandating a coherent approach that begins with stage-aware problem definition and flows through data synthesis, principled feature engineering, rigorous validation, and risk-aware interpretation, the SAISE framework offers a new standard for conducting research in this domain.

Ultimately, this work serves as a call for a more mature and disciplined approach to AI-driven startup evaluation. As these technologies move from the lab to live investment decisions, the need for models that are not only accurate but also robust, transparent, and theoretically grounded becomes paramount. By embracing a more systematic methodology that augments, rather than replaces, human judgment (Shepherd & Majchrzak, 2022; Dellermann et al., 2017) and fosters a true "human-AI symbiosis" (Jarrahi, 2018), researchers can accelerate the field's progress, moving beyond simple prediction to create the next generation of AI tools capable of delivering genuine, trustworthy insight into the complex phenomenon of startup success.

# Appendix

## Appendix A: Full Table of Predictor Feature Families

### Table A1: Thematic Landscape and Prevalence of Predictor Feature Families (Exhaustive List, N=57)

This table provides a hierarchical overview of the feature families identified across the 57-study corpus, ranked by their frequency of use. It highlights the field's strong consensus on a core "money-people-market" triad while also cataloging the long tail of more specialized and emerging indicators.

| Rank | Feature Family | Core Concept & Representative Variables | Prevalence (N=57) | All Studies in Category |
|---|---|---|---|---|
| 1 | Funding History & Cadence | The magnitude, timing, and velocity of capital acquisition. Total funding (USD), # of rounds, time-to-first-round, gap-between-rounds. | 39 (68%) | Abhinand & Poonam (2022), Adebiyi et al. (2024), Antretter et al. (2018), Belgaum et al. (2024), Chen. (2024), Cholil et al. (2024), Choi (2024), Deodhar et al. (2024), Dellermann et al. (2017), Dziubanovska et al. (2024), Eljil & Nait-Abdesselam (2024), Espinoza-Mina & Colina Vargas (2024), Ferrati et al. (2021), Ferry et al. (2018), Font-Cot et al. (2025), Gangwani et al. (2023), Goossens et al. (2023), Gracy et al. (2024), Guerzoni et al. (2021), Kim et al. (2023), Kalbande & Karmore (2024), Li, Y. et al. (2024), Li, J. (2020), Liu et al. (2022), Misra, Jat, & Mishra (2023), Pasayat & Bhowmick (2021), Pasayat & Bhowmick (2023), Razaghzadeh Bidgoli et al. (2024), Ross et al. (2021), Samudra & Satya (2024), Setty et al. (2024), Sharchilev et al. (2018), Shi et al. (2024), Srivani et al. (2023), Tagkouta et al. (2023), Te, Liu, & Müller (2023), Ungerer et al. (2021), William et al. (2022), Żbikowski & Antosiuk (2021) |
| 2 | Team / Founder Attributes | The human capital, experience, and composition of the founding team. Founder count, prior exits, education pedigree, skill diversity. | 34 (60%) | Abhinand & Poonam (2022), Al Rahma & Abrar-Ul-Haq (2024), Antretter et al. (2018), Belgaum et al. (2024), Cholil et al. (2024), Deodhar et al. (2024), Dellermann et al. (2017), Eljil & Nait-Abdesselam (2024), Ferry et al. (2018), Font-Cot et al. (2025), Gangwani et al. (2023), Goossens et al. (2023), Gracy et al. (2024), Guerzoni et al. (2021), Kalbande & Karmore (2024), Kim et al. (2023), Li, Y. et al. (2024), Li, J. (2020), Liu et al. (2022), Misra, Jat, & Mishra (2023), Neugebauer et al. (2021), Pasayat & Bhowmick (2021), Pasayat & Bhowmick (2023), Rawat et al. (2025), Ross et al. (2021), Sabahi & Parast (2020), Schade & Schuhmacher (2023), Setty et al. (2024), Shi et al. (2024), Srivani et al. (2023), Takas et al. (2024), Te, Liu, & Müller (2023), William et al. (2022), Żbikowski & Antosiuk (2021) |
| 3 | Market / Sector Tags | The industrial and competitive context in which the startup operates. One-hot industry codes, BERT topic vectors, | 34 (60%) | Abhinand & Poonam (2022), Belgaum et al. (2024), Chen. (2024), Cholil et al. (2024), Choi (2024), Dellermann et al. (2017), Dziubanovska et al. (2024), Eljil & Nait-Abdesselam (2024), Espinoza-Mina & Colina Vargas (2024), Font-Cot et al. (2025), Gangwani et al. (2023), Goossens et al. (2023), Gracy et al. (2024), Guerzoni et al. |
| | | market growth dummies. | | (2021), Kalbande & Karmore (2024), Kim et al. (2023), Li, Y. et al. (2024), Li, X. et al. (2024), Misra, Jat, & Mishra (2023), Pasayat & Bhowmick (2021), Rawat et al. (2025), Ross et al. (2021), Samudra & Satya (2024), Schade & Schuhmacher (2023), Setty et al. (2024), Sharchilev et al. (2018), Shi et al. (2024), Srivani et al. (2023), Tagkouta et al. (2023), Takas et al. (2024), Te, Liu, & Müller (2023), Ungerer et al. (2021), William et al. (2022), Żbikowski & Antosiuk (2021) |
| 4 | Investor Structure / Quality | The quality and composition of the startup's financial backers. Investor count, lead/brand flags, top-tier investor dummy. | 26 (46%) | Abhinand & Poonam (2022), Belgaum et al. (2024), Chen. (2024), Cholil et al. (2024), Deodhar et al. (2024), Dellermann et al. (2017), Dziubanovska et al. (2024), Eljil & Nait-Abdesselam (2024), Ferrati et al. (2021), Gangwani et al. (2023), Gracy et al. (2024), Guerzoni et al. (2021), Kalbande & Karmore (2024), Kim et al. (2023), Li, Y. et al. (2024), Li, X. et al. (2024), Misra, Jat, & Mishra (2023), Pasayat & Bhowmick (2021), Ross et al. (2021), Samudra & Satya (2024), Setty et al. (2024), Sharchilev et al. (2018), Shi et al. (2024), Srivani et al. (2023), Te, Liu, & Müller (2023), Żbikowski & Antosiuk (2021) |
| 5 | Digital & Social Traction | Proxies for market presence, brand momentum, and user engagement. Website liveness, LinkedIn/Twitter followers, web mention freshness. | 20 (35%) | Antretter et al. (2018), Belgaum et al. (2024), Cholil et al. (2024), Eljil & Nait-Abdesselam (2024), Goossens et al. (2023), Kim et al. (2023), Li, Y. et al. (2024), Misra, Jat, & Mishra (2023), Pasayat & Bhowmick (2021), Razaghzadeh Bidgoli et al. (2024), Ross et al. (2021), Samudra & Satya (2024), Setty et al. (2024), Sharchilev et al. (2018), Tagkouta et al. (2023), Takas et al. (2024), Te, Liu, & Müller (2023), Ungerer et al. (2021), William et al. (2022), Żbikowski & Antosiuk (2021) |
| 6 | Product / Technology Sophistication | The maturity and defensibility of the startup's core technology or product. Patent counts, Technology Readiness Level (TRL), GitHub velocity. | 14 (24%) | Dellermann et al. (2017), Ferrati et al. (2021), Goossens et al. (2023), Kim et al. (2023), Li, Y. et al. (2024), Misra, Jat, & Mishra (2023), Pasayat & Bhowmick (2023), Rawat et al. (2025), Ross et al. (2021), Samudra & Satya (2024), Takas et al. (2024), Ungerer et al. (2021), William et al. (2022), Żbikowski & Antosiuk (2021) |
| 7 | Financial-Statement Ratios | Traditional accounting metrics reflecting a firm's financial health. Cash/asset ratio, debt-to-cash-flow, EBITDA margin, burn-rate. | 8 (14%) | Adebiyi et al. (2024), Dellermann et al. (2017), Font-Cot et al. (2025), Guerzoni et al. (2021), Liu et al. (2022), Misra, Jat, & Mishra (2023), Samudra & Satya (2024), Takas et al. (2024) |
| 8 | Sentiment & Emotion | Measures of public or investor affect derived from textual data. VADER polarity, RoBERTa emotion vectors, tweet sentiment. | 3 (5%) | Antretter et al. (2018), Goossens et al. (2023), Sharchilev et al. (2018) |
| 9 | Psychometric / Behavioural | Direct measures of founder personality and behavioral traits. Big-Five OCEAN scores, need-for-achievement, resilience. | 4 (7%) | Al Rahma & Abrar-Ul-Haq (2024), Font-Cot et al. (2025), Sabahi & Parast (2020), Takas et al. (2024) |
| 10 | ESG & Sustainability Cues | Indicators reflecting a startup's focus on environmental or social goals. E/S/G keyword dummies, green-impact flags. | 2 (3%) | Rani et al. (2024), Takas et al. (2024) |
| 11 | Novel Engineered / | Composite features modeling the interplay between different | 3 (5%) | Gangwani et al. (2023), Li, Y. et al. (2024), Ross et al. (2021) |

| | | | | | |
|---|---|---|---|---|---|
| Interaction Metrics | | entities. Investor-Market-Fit, Founder-Idea-Fit cosine. | | | |

## Appendix B: Full Table of Target Variable Operationalizations

### Table B1: Thematic Landscape of Target Variable Operationalizations for "Startup Success" (Exhaustive List, N=57)

This table deconstructs the various ways "startup success" is defined and measured across the reviewed literature (N=57), grouping them into six conceptual themes. It highlights the critical fragmentation of the ground truth—from funding milestones and exit events to long-term survival—which has profound implications for the comparability and interpretation of model performance across studies.

| Conceptual Theme | Prevalence (N=57) | Operationalization & Representative Examples | Typical Horizon | Core Rationale / Purpose | All Studies in Category |
|---|---|---|---|---|---|
| **Exit Events / Firm Status** | 19 (33%) | Success is a firm's ultimate liquidity event or operational status, using binary or multi-class labels. • IPO or M&A (binary exit) • Operating/Acquired/IPO/Closed. | Snapshot to 10 years | Captures the canonical goal of venture capital by identifying a definitive realization of success or failure. | Kim et al. (2023), Setty et al. (2024), Ferrati et al. (2021), Ross et al. (2021), Gautam & Wattanapongsakorn (2024), Misra, Jat, & Mishra (2023), Mishra, Jat, & Mishra (2023), Cholil et al. (2024), Choi (2024), Samudra & Satya (2024), Gangwani et al. (2023), Srivani et al. (2023), Gracy et al. (2024), Guerzoni et al. (2021), Pasayat & Bhowmick (2021), Pasayat & Bhowmick (2023), Ungerer et al. (2021), Hsairi (2024), Vasquez et al. (2023) |
| **Funding Milestone** | 13 (23%) | Success is a capital acquisition event, typically an institutional VC round. • Raises a first Series-A round • Secures Series-B funding. | 1-5 years | Tracks whether the venture succeeds in attracting external capital, signaling early market validation and investor confidence. | Te, Liu, & Müller (2023), Dellermann et al. (2017), Żbikowski & Antosiuk (2021), Li, Y. et al. (2024), Razaghzadeh Bidgoli et al. (2024), Belgaum et al. (2024), Sharchilev et al. (2018), Eljil & Nait-Abdesselam (2024), Kalbande & Karmore (2024), Li J. (2020), Shi et al. (2024), Tagkouta et al. (2023), Te, Wieland, Frey, et al. (2023) |
| **Survival, Lifespan & Sustainability** | 6 (11%) | Success is long-term viability, moving beyond discrete events to capture continuity and resilience. • 5-year survival probability • Expert-rated sustainability score. | 5 years to full lifespan | Measures the firm's capacity to operate viably over time, often incorporating non-financial aspects. | Antretter et al. (2018), Espinoza-Mina & Colina Vargas (2024), Font-Cot et al. (2025), Takas et al. (2024), Wang et al. (2022), Park et al. (2024) |
| **Entrepreneurial / Project Action** | 6 (11%) | Success is defined by the completion of an entrepreneurial action or project, distinct from firm-level outcomes. • Predicting investor decisions • Kickstarter | Short-term | Focuses on predicting intermediate actions or project- | Goossens et al. (2023), Schade & Schuhmacher (2023), Sabahi & Parast (2020), Dziubanovska et al. (2024), Deodhar et al. (2024), Rawat et al. (2025) |

| | | | | |
|---|---|---|---|---|
| | | campaign success • Shark Tank deal prediction. | | level success rather than the venture's ultimate fate. |
| **Valuation & Investor Return** | 4 (7%) | Success is quantified via direct economic value or achieving a landmark valuation. • Market-cap regression • Unicorn status (≥ $1B valuation). | Snapshot to full fund life | Quantifies the economic upside for investors, moving beyond simple binary success. | Abhinand & Poonam (2022), Chen. (2024), Hu et al. (2022), Zhang et al. (2023) |
| **Financial Distress / Health** | 2 (4%) | Success is framed as the avoidance of failure or near-term solvency. • Bankruptcy or acquisition within 12 months. | 1-year projection | Provides early-warning signals of impending insolvency, allowing for pre-emptive intervention. | Adebiyi et al. (2024), Liu et al. (2022) |

Note: The N=57 represents the total count of distinct empirical studies in the corpus. Each study is categorized under one primary target variable theme based on its main empirical model.

## Appendix C: Full Table of AI/ML Algorithm Families and Their Application

### Table C1: Frequency and Application of AI/ML Algorithm Families Across the Corpus (Exhaustive List, N=57)

This table presents a quantitative breakdown of the modeling techniques used across the 57-paper corpus, organized by algorithmic family. The final column provides an exhaustive list of citations for each family, offering a detailed, traceable map of the field's methodological landscape.

| Rank | Algorithm Family | Core Concept & Representative Algorithms | Prevalence (N=57) | All Studies Employing This Family (Citations) |
|---|---|---|---|---|
| 1 | Tree-based | Algorithms that create hierarchical, rule-based models. • Decision Tree (CART), Random Forest, Extra-Trees | 39 (72%) | Abhinand & Poonam (2022), Antretter et al. (2018), Belgaum et al. (2024), Cholil et al. (2024), Deodhar et al. (2024), Dellermann et al. (2017), Dziubanovska et al. (2024), Espinoza-Mina & Colina Vargas (2024), Ferrati et al. (2021), Font-Cot et al. (2025), Gangwani et al. (2023), Gautam & Wattanapongsakorn (2024), Goossens et al. (2023), Gracy et al. (2024), Guerzoni et al. (2021), Hsairi (2024), Hu et al. (2022), Kalbande & Karmore (2024), Kim et al. (2023), Li, Y. et al. (2024), Li, J. (2020), Liu et al. (2022), Misra, Jat, & Mishra (2023), Pasayat & Bhowmick (2021), Pasayat & Bhowmick (2023), Razaghzadeh Bidgoli et al. (2024), Ross et al. (2021), Sabahi & Parast (2020), Samudra & Satya (2024), Setty et al. (2024), Sharchilev et al. (2018), Shi et al. (2024), Srivani et al. (2023), Tagkouta et al. (2023), Takas et al. (2023), Te, Liu, & Müller (2023), Vasquez et al. (2023), William et al. (2022), Żbikowski & Antosiuk (2021) |
| 2 | Ensemble / | Combining multiple models to | 37 (67%) | Abhinand & Poonam (2022), Antretter et al. (2018), Belgaum et al. (2024), Chen. (2024), |
| | Boosting | improve predictive performance. • Gradient Boosting, XGBoost, LightGBM, AdaBoost, CatBoost, Stacking | | Cholil et al. (2024), Deodhar et al. (2024), Dellermann et al. (2017), Espinoza-Mina & Colina Vargas (2024), Ferrati et al. (2021), Font-Cot et al. (2025), Gangwani et al. (2023), Gautam & Wattanapongsakorn (2024), Goossens et al. (2023), Gracy et al. (2024), Hsairi (2024), Hu et al. (2022), Kalbande & Karmore (2024), Kim et al. (2023), Li, Y. et al. (2024), Li, J. (2020), Liu et al. (2022), Misra, Jat, & Mishra (2023), Pasayat & Bhowmick (2021), Pasayat & Bhowmick (2023), Razaghzadeh Bidgoli et al. (2024), Ross et al. (2021), Samudra & Satya (2024), Schade & Schuhmacher (2023), Setty et al. (2024), Sharchilev et al. (2018), Shi et al. (2024), Srivani et al. (2023), Tagkouta et al. (2023), Takas et al. (2024), Te, Liu, & Müller (2023), Vasquez et al. (2023), William et al. (2022) |
| 3 | Linear / GLM | Models assuming a linear relationship between features and the target. • Logistic/Linear Regression, Ridge/Lasso, Elastic-net | 29 (54%) | Adebiyi et al. (2024), Antretter et al. (2018), Dellermann et al. (2017), Font-Cot et al. (2025), Gangwani et al. (2023), Gautam & Wattanapongsakorn (2024), Gracy et al. (2024), Guerzoni et al. (2021), Gujarathi et al. (2024), Hsairi (2024), Hu et al. (2022), Kalbande & Karmore (2024), Kim et al. (2023), Li, Y. et al. (2024), Li, J. (2020), Liu et al. (2022), Misra, Jat, & Mishra (2023), Pasayat & Bhowmick (2023), Razaghzadeh Bidgoli et al. (2024), Ross et al. (2021), Sabahi & Parast (2020), Schade & Schuhmacher (2023), Sharchilev et al. (2018), Srivani et al. (2023), Tagkouta et al. (2023), Takas et al. (2024), Vasquez et al. (2023), William et al. (2022), Żbikowski & Antosiuk (2021) |
| 4 | Kernel / Instance-based | Models making predictions based on feature similarity or distance. • Support Vector Machine (SVM), k-Nearest Neighbors (KNN) | 29 (53%) | Abhinand & Poonam (2022), Antretter et al. (2018), Cholil et al. (2024), Dellermann et al. (2017), Gangwani et al. (2023), Gracy et al. (2024), Hsairi (2024), Hu et al. (2022), Kalbande & Karmore (2024), Kim et al. (2023), Li, Y. et al. (2024), Li, J. (2020), Liu et al. (2022), Misra, Jat, & Mishra (2023), Pasayat & Bhowmick (2023), Razaghzadeh Bidgoli et al. (2024), Ross et al. (2021), Sabahi & Parast (2020), Schade & Schuhmacher (2023), Setty et al. (2024), Sharchilev et al. (2018), Shi et al. (2024), Srivani et al. (2023), Tagkouta et al. (2023), Takas et al. (2024), Vasquez et al. (2023), William et al. (2022), Żbikowski & Antosiuk (2021) |
| 5 | Neural Networks / Deep Learning | Biologically-inspired models with interconnected layers of nodes. • MLP/ANN, CNN, Bi-LSTM, TextCNN, LLM-agents | 19 (35%) | Abhinand & Poonam (2022), Antretter et al. (2018), Chen. (2024), Dellermann et al. (2017), Ferrati et al. (2021), Font-Cot et al. (2025), Gangwani et al. (2023), Goossens et al. (2023), Gracy et al. (2024), Hsairi (2024), Liu et al. (2022), Misra, Jat, & Mishra (2023), Pasayat & Bhowmick (2023), Rawat et al. (2025), Ross et al. (2021), Sabahi & Parast (2020), Srivani et al. (2023), Takas et al. (2024), Te, Liu, & Müller (2023) |
| 6 | Bayesian / Probabilistic | Models based on applying Bayes' theorem and probability distributions. • Naïve Bayes, Bayesian Networks | 9 (16%) | Al Rahma & Abrar-Ul-Haq (2024), Abhinand & Poonam (2022), Cholil et al. (2024), Dellermann et al. (2017), Gujarathi et al. (2024), Kim et al. (2023), Misra, Jat, & Mishra (2023), Schade & Schuhmacher (2023), Takas et al. (2024) |

Note: Prevalence counts indicate the number of unique studies in the N=57 corpus employing at least one algorithm from that family. As studies often compare multiple algorithms, the shares are not mutually exclusive and sum to more than 100%. The citation lists are comprehensive based on our final inclusion criteria.

**Appendix D: Full Table of Data Source Families and Their Application**

**Table D1: Taxonomy and Prevalence of Data Source Families in Startup Prediction (Exhaustive List, N=57)**

This table provides a classification of the distinct raw data sources identified across the 57-study corpus, organized into nine families. It quantifies the field's bifurcated data strategy and provides a complete list of citations for each family.

| Data Source Family | Representative Raw Sources | Paper Count | Distinct Sources | All Studies Employing Family (Citations) |
|---|---|---|---|---|
| Venture Databases | Crunchbase, AngelList, Seed-DB, IT Orange, Zero2IPO, Dealroom, Mattermark, CB Insights | 28 (49%) | 8 | Belgaum et al. (2024), Chen. (2024), Choi (2024), Cholil et al. (2024), Dellermann et al. (2017), Ferrati et al. (2021), Gangwani et al. (2023), Gautam & Wattanapongsakorn (2024), Gracy et al. (2024), Guerzoni et al. (2021), Kim et al. (2023), Li, Y. et al. (2024), Misra, Jat, & Mishra (2023), Pasayat & Bhowmick (2021), Pasayat & Bhowmick (2023), Razaghzadeh Bidgoli et al. (2024), Ross et al. (2021), Samudra & Satya (2024), Schade & Schuhmacher (2023), Setty et al. (2024), Sharchilev et al. (2018), Shi et al. (2024), Srivani et al. (2023), Te, Liu, & Müller (2023), Ungerer et al. (2021), Wang et al. (2022), William et al. (2022), Żbikowski & Antosiuk (2021) |
| Social-media / Open-web | Twitter, Facebook, LinkedIn, GitHub Stars, Telegram, Medium, BitcoinTalk, Yandex Web-graph, Google Search hits, Alexa Web Traffic | 20 (35%) | 11 | Antretter et al. (2018), Belgaum et al. (2024), Cholil et al. (2024), Eljil & Nait-Abdesselam (2024), Goossens et al. (2023), Kim et al. (2023), Li, Y. et al. (2024), Misra, Jat, & Mishra (2023), Pasayat & Bhowmick (2021), Razaghzadeh Bidgoli et al. (2024), Ross et al. (2021), Samudra & Satya (2024), Setty et al. (2024), Sharchilev et al. (2018), Tagkouta et al. (2023), Takas et al. (2024), Te, Liu, & Müller (2023), Ungerer et al. (2021), William et al. (2022), Żbikowski & Antosiuk (2021) |
| Survey / Lab / Expert Panel | Founder/Sustainability Surveys, Incubator/Expert Scores, Investor Crowd Ratings, Bahrain EO Survey, Nigerian E-commerce Financials | 8 (14%) | 6 | Adebiyi et al. (2024), Al Rahma & Abrar-Ul-Haq (2024), Dellermann et al. (2017), Font-Cot et al. (2025), Gujarathi et al. (2024), Sabahi & Parast (2020), Tagkouta et al. (2023), Takas et al. (2024) |
| Registry / Filings | USPTO PatentsView, SEC EDGAR Filings, EU VICO 2.0 Balance Sheets, National Company Registries | 5 (9%) | 4 | Chen. (2024), Ferrati et al. (2021), Kim et al. (2023), Liu et al. (2022), Ross et al. (2021) |
| Academic Benchmark Set | Kaggle "DPhi Startup", Kaggle "Startup Success", Kaggle "Software-Startups", Public Crunchbase CSVs | 4 (7%) | 4 | Choi (2024), Cholil et al. (2024), Gracy et al. (2024), Samudra & Satya (2024) |
| Hybrid Scrape / Composites | Venture Intelligence, StartupTalky, TechCrunch (News), Alexa Web-traffic (as part of a multi-scrape) | 4 (7%) | 4 | Abhinand & Poonam (2022), Belgaum et al. (2024), Dellermann et al. (2017), Razaghzadeh Bidgoli et al. (2024) |
| Token / Crowdfunding Platforms | Kickstarter, Wadiz, Tumblbug, Crowdy, ICOBench, ICOdrops | 2 (4%) | 6 | Dziubanovska et al. (2024), Eljil & Nait-Abdesselam (2024) |
| Search-API Enrichments | Google Custom-Search API | 1 (2%) | 1 | Samudra & Satya (2024) |
| Document | Pitch-deck PDFs, Excel Financials | 1 (2%) | 2 | Samudra & Satya (2024) |

| Upload / Internal Files | |
|---|---|

Note: The sum of the "Paper Count" column exceeds the total corpus size of 57 because many studies employ a multi-modal data strategy, drawing from more than one source family. "Distinct Sources" refers to the number of unique, named raw data sources identified within each family across the entire corpus.